\documentclass[10pt]{article}
\usepackage{IEEEtrantools}
\usepackage{cite}

   \usepackage[dvips]{graphicx}
   \graphicspath{{./eps/}}
   \DeclareGraphicsExtensions{.eps}
\usepackage[caption=false,font=footnotesize]{subfig}
\usepackage{stfloats}
\begin{document}
%
\title{Motion-Compensated Coding and\\Frame-Rate Up-Conversion: Models and Analysis}
%
%
%

\author{Yehuda~Dar
        and~Alfred~M.~Bruckstein
\\\\Technion -- Israel Institute of Technology\\Haifa 32000, Israel
\\\\E-mail: ydar@tx.technion.ac.il, freddy@cs.technion.ac.il
}

%
%

\markboth{Dar and Bruckstein: Motion-Compensated Coding and Frame-Rate Up-Conversion}%
{Dar and Bruckstein: Motion-Compensated Coding and Frame-Rate Up-Conversion}
%



\date{}

\maketitle

\begin{abstract}

Block-based motion estimation (ME) and compensation (MC) techniques are widely used in modern video processing algorithms and compression systems. The great variety of video applications and devices results in numerous compression specifications. Specifically, there is a diversity of frame-rates and bit-rates. 
In this paper, we study the effect of frame-rate and compression bit-rate on block-based ME and MC as commonly utilized in inter-frame coding and frame-rate up conversion (FRUC).
This joint examination yields a comprehensive foundation for comparing MC procedures in coding and FRUC.
First, the video signal is modeled as a noisy translational motion of an image. Then, we theoretically model the motion-compensated prediction of an available and absent frames as in coding and FRUC applications, respectively. The theoretic MC-prediction error is further analyzed and its autocorrelation function is calculated for coding and FRUC applications.
We show a linear relation between the variance of the MC-prediction error and temporal-distance. 
While the affecting distance in MC-coding is between the predicted and reference frames, MC-FRUC is affected by the distance between the available frames used for the interpolation.
Moreover, the dependency in temporal-distance implies an inverse effect of the frame-rate.
FRUC performance analysis considers the prediction error variance, since it equals to the mean-squared-error of the interpolation. However, MC-coding analysis requires the entire autocorrelation function of the error; hence, analytic simplicity is beneficial. Therefore, we propose two constructions of a separable autocorrelation function for prediction error in MC-coding.
We conclude by comparing our estimations with experimental results.
\end{abstract}



%

\section{Introduction}
%
%
%
%
Temporal redundancy is a main property of video signals. This redundancy originates in the similarity between successive frames in a video scene. Moreover, a video scene can be thought of as a composition of static and moving  regions. Therefore, many video compression and processing systems utilize motion estimation (ME). Ideally, the motion should be estimated per pixel; however, practical systems have run-time limitations, and therefore cannot apply estimation per pixel. Hence, block-based ME techniques are widely used in practical video compression and processing algorithms. Block-based ME is the procedure of estimating block motion by comparing it with blocks in a search area within another frame in the sequence. This method approximates the motion as translational, and represents it by a motion-vector and reference frame indication.

In block-based hybrid compression, ME is utilized for inter-frame prediction of a coded block. Motion-compensation (MC) is the subtraction between the coded block and its prediction. This results in the block's prediction error, also known as the MC-residual. The MC-residual is further coded and sent to the decoder. Therefore, the MC-residual greatly affects the performance of inter-frame coding. Furthermore, due to the extensive use of inter-frame coding, the MC-residual also significantly influences the overall compression performance.

Accordingly, the MC-residual has been widely studied since the 1980's \cite{RefWorks:4, RefWorks:5, RefWorks:6, RefWorks:11, RefWorks:10, RefWorks:8, RefWorks:7, RefWorks:12, RefWorks:9, RefWorks:57, RefWorks:60, RefWorks:61}. However, these studies have not explicitly considered the frame-rate effect on the MC-residual statistics. Moreover, only Guo et al. \cite{RefWorks:57} mentioned the influence of frame-reconstruction quality (and, therefore, bit-rate) on the MC-residual. In this paper, we analyze the effects of frame-rate (through the temporal-distance) and bit-rate on the MC-residual autocorrelation function. Most of the available analytic models are too complex for being a basis for analysis of an entire compression system. Here we propose two models for the autocorrelation. First, we derive a rather complex expression from our theoretic model for MC-prediction of an available frame (i.e., as in coding). Then, we simplify the autocorrelation to a separable form similar to \cite{RefWorks:60} and \cite{RefWorks:10}. Furthermore, we justify our analysis by experimental observations.

Frame-rate up conversion (FRUC) is the procedure of increasing the frame-rate of a video by temporal interpolation of frames. There are several motivations for using FRUC. It is used for video format conversion when the target format has higher frame-rate. In addition, high frame-rates were found to increase the subjective quality \cite{RefWorks:42}; therefore, some applications may apply FRUC on low frame-rate videos. Another application of FRUC is for improving low bit-rate video coding as follows: the frame-rate is reduced before compression, and increased back to its original value after the reconstruction of the compressed data. As a result, the output video quality is improved for a constant bit-budget.
	
FRUC algorithms trade off between computational complexity and the quality of the interpolated frames. Simple FRUC techniques disregard the motion in the sequence, e.g., interpolating by frame repetition or averaging. For non-static regions, this results in motion jerkiness and ghost artifacts. Therefore, the commonly used interpolation techniques consider motion.	Specifically, methods that utilize motion-trajectory estimation are known as motion-compensated FRUC (MC-FRUC).

Some studies have proposed complex FRUC algorithms that try to accurately model the motion in the video, e.g., \cite{RefWorks:45}. However, high computational complexity limits these algorithms for offline usage, whereas some applications require real-time FRUC. A reasonable computational complexity is achieved in block-based MC-FRUC techniques; therefore, they are widely used and studied \cite{RefWorks:46, RefWorks:47, RefWorks:39, RefWorks:41, RefWorks:49}. 
Block-based MC-FRUC is usually performed by applying block-matching procedure between existing frames, resulting in a trajectory of the estimated translational motion; then,
this motion-trajectory is used for interpolating missing blocks according to the applied method \cite{RefWorks:46, RefWorks:47, RefWorks:39, RefWorks:41, RefWorks:49}.


In \cite{RefWorks:49}, the MC-FRUC error was analyzed in the power-spectral-density (PSD) domain and by using a statistical model of the motion-vector error. They searched for the optimal temporal filter. In this paper, we study the block-based MC-FRUC error in the pixel domain. 
The examined procedure models low-complexity methods (e.g., \cite{RefWorks:46}), which are commonly used. Consequently, the proposed analytic derivations are relatively simple.

Block-based ME differs from the true motion by assuming it is translational. This sub-optimality has minor importance in the application of MC for inter-frame coding, where the motion estimation is performed at the encoder between two accessible frames, and the target is minimal prediction residual. However, ME in FRUC aims at estimating the true motion in a missing frame. Therefore, the translational motion assumption deteriorates MC-FRUC performance. Dane and Nguyen \cite{RefWorks:49} discussed the differences between the application of MC to coding and FRUC. This paper continues this examination by giving side by side analyses of MC-coding and MC-FRUC, which are easily comparable due to joint assumptions and mathematical tools.

This paper is organized as follows. 
In section \ref{sec:Video Signal Model}, we present a theoretic model for the video signal.
Section \ref{sec:Analysis of Motion-Compensated Prediction} analyzes the MC-prediction and its error for the cases of available and absent frames, i.e., coding and FRUC, respectively.
Section \ref{sec:Simplified Autocorrelation Model} introduces two constructions of a separable autocorrelation function for MC-coding. In section \ref{sec:Theoretical Estimations} we study the theoretic estimations of our model.
In section \ref{sec:Experimental Results} we present experimental results to validate our models. 
Section \ref{sec:Conclusion} concludes this paper.

\section{Video Signal Model}
\label{sec:Video Signal Model}
\subsection{A Noised Translational Motion Model}
The digital video signal is a temporal sequence of 2D images, i.e. $\left\{ {{f_t}\left( {x,y} \right)} \right\}_{t = 0}^{T}$. Adjacent frames are known to be correlated; hence, we relate the frames by assuming a translational motion of a 2D image with additive noise process.

We assume that the frame sequence $\left\{ {{f_t}\left( {x,y} \right)} \right\}_{t = 0}^{T}$ is decomposable into two sequences. First, a 2D image with a translational motion denoted as $\left\{ {{v_t}\left( {x,y} \right)} \right\}_{t = 0}^{T}$. Second, a temporally-accumulated noise process, $\left\{ {{n_t}\left( {x,y} \right)} \right\}_{t = 0}^{T}$, that represents differences between $\left\{ {{v_t}\left( {x,y} \right)} \right\}_{t = 0}^{T}$ and the actual frames due to deviations from translational motion such as deformations of objects, camera noise and quantization noise. The proposed decomposition is expressed as follows.
\begin{equation}
\label{eq:frame decomposition equation}
{f_t}\left( {x,y} \right) = {v_t}\left( {x,y} \right) + {n_t}\left( {x,y} \right)
\end{equation}

The underlying translational motion process is defined as follows. The motion at the $t^{th}$ frame relative to its predecessor at $t - 1$ is denoted as $\varphi \left( {t,t - 1} \right) = \left( {{\varphi _x}\left( {t,t - 1} \right),{\varphi _y}\left( {t,t - 1} \right)} \right)$. Hence, the motion in the video can be represented by the sequence $\left\{ {\varphi \left( {i,i - 1} \right)} \right\}_{i = 1}^{T}$. Moreover, the motion between two time points, $t_1$ and $t_2$, is defined as follows.
\begin{equation}
\label{eq:motion between two frames}
\varphi \left( {{t_2},{t_1}} \right) = \left\{ \begin{array}{l}
\sum\limits_{i = {t_1} + 1}^{{t_2}} {\varphi \left( {i,i - 1} \right)} \,\,\,\,\,\,\,\,\,\,\,\,\,\,\,\,,\,\,\,\,for\,\,\,\,{t_1} < {t_2}\\
 - \sum\limits_{i = {t_2} + 1}^{{t_1}} {\varphi \left( {i,i - 1} \right)} \,\,\,\,\,\,\,\,\,\,\,\,\,,\,\,\,\,for\,\,\,\,{t_2} < {t_1}\\
\left( {0,0} \right)\,\,\,\,\,\,\,\,\,\,\,\,\,\,\,\,\,\,\,\,\,\,\,\,\,\,\,\,\,\,\,\,\,\,,\,\,\,\,for\,\,\,\,{t_1} = {t_2}
\end{array} \right.
\end{equation}

We model $v_t$ to be a constant base frame, $v$, spatially shifted by $\left( {{\varphi _x}\left( {t,0} \right),{\varphi _y}\left( {t,0} \right)} \right)$, i.e.,
\begin{equation}
\label{eq:v_t definition}
{v_t}\left( {x,y} \right) = v\left( {x - {\varphi _x}\left( {t,0} \right),y - {\varphi _y}\left( {t,0} \right)} \right)
\end{equation}
The image $v$ is assumed to be wide-sense stationary (WSS) and is modeled using first-order Markov process, i.e., its autocorrelation is 
\begin{equation}
\label{eq:pure image autocorrelation}
{R_v}\left( {k,l} \right) = \sigma _v^2\cdot{\rho _v}^{\left| k \right| + \left| l \right|}.
\end{equation}

\begin{figure}
\centering
\includegraphics[height=3.5in]{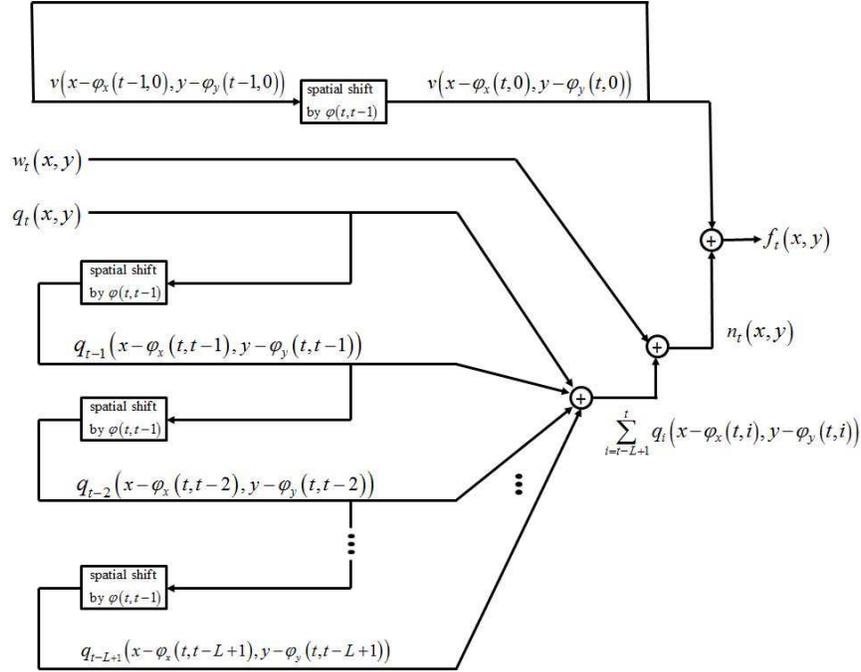}
\caption{Demonstration of the proposed video model.}
\label{Fig:Video_model}
\end{figure}

We model the noise, $n_t$, as a combination of two elements. 
Firstly, a temporally-local noise, $w_t$, that represents distortions that are relevant only for the frame at time $t$, e.g., camera noise or quantization noise.
Secondly, we represent object deformations using a temporally-accumulated noise process.
We assume that frames have equal average energy (i.e., $f_t$ has a constant variance for any $t$). Therefore, there is a fixed amount of object deformation relative to the original form in $v$; otherwise, the immersion of $v$ in noise will increase over time. Consequently, the noise component, $n_t$, also represents the accumulated deviation from translational motion along the recent $L$ frames; i.e., the process has finite memory of length $L$. For $n_t$'s construction, we use an auxiliary noise sequence $\left\{ {{q_t}} \right\}_{t =  - L + 1}^\infty $, which is a spatially i.i.d random variable with a zero-mean Gaussian distribution with variance $\sigma _{q}^2$. Moreover, ${q_i}$ is independent from ${q_k}$ for $i \ne k$, and from $w_j$ for any $j$.
We assume that at each time point, $t$, spatial noise signals $q_t$ and $w_t$ are introduced and affect $n_t$ together with the last $L - 1$ preceding $q_k$ elements (Fig. \ref{Fig:Video_model}), i.e.
\begin{equation}
\label{eq:accumulated noise definition}
{n_t}\left( {x,y} \right) = {w_t}\left( {x,y} \right) + \sum\limits_{i = t - {L} + 1}^t {{q_i}\left( {x - {\varphi _x}\left( {t,i} \right),y - {\varphi _y}\left( {t,i} \right)} \right)}. 
\end{equation}
Where we utilized the property $\varphi \left( {t,t} \right) = \left( {0,0} \right)$.
Recall that $q_i$ is available also for negative time points starting at $t = - {L} + 1$. Consequently, the temporally-accumulated noise has a spatially i.i.d, zero-mean Gaussian distribution with variance $L \cdot \sigma _q^2$ for any $t$. Accordingly, ${n_t}\left( {x,y} \right)$'s autocorrelation is 
\begin{equation}
\label{eq:accumulated noise autocorrelation}
{R_{{n_t}}}\left( {k,l} \right) = \left( {\sigma _w^2 + L\sigma _q^2} \right)\cdot\delta \left( {k,l} \right)
\end{equation}

Setting (\ref{eq:v_t definition}) and (\ref{eq:accumulated noise definition}) into (\ref{eq:frame decomposition equation}) yields
\begin{IEEEeqnarray}{rCl}
\label{eq:frame decomposition with translational MV}
{f_t}\left( {x,y} \right) & = & v\left( {x - {\varphi _x}\left( {t,0} \right),y - {\varphi _y}\left( {t,0} \right)} \right) 
\\ \nonumber
&& + {w_t}\left( {x,y} \right) + \sum\limits_{i = t - {L} + 1}^{t} {{q_i}\left( {x - {\varphi _x}\left( {t,i} \right),y - {\varphi _y}\left( {t,i} \right)} \right)}.
\end{IEEEeqnarray}

\subsection{Frame-Rate Effect}
The variance of the auxiliary noise elements, $\sigma_q^2$, reflects the energy of the differences between successive frames that cannot be perfectly estimated by a translational transformation, even for a continuous images (i.e., the estimation algorithm has no spatial accuracy issues). We point here on two affecting factors: frame-rate and the compression bit-rate.

The frame-rate, $F_{rate}$, defines the time-intervals between successive frames to be $\frac{1}{F_{rate}}$. We assume that the energy of the modifications expressed in $\sigma_q^2$ is linear in the temporal-distance. Hence, 
\begin{IEEEeqnarray}{rCl}
\label{eq:Difference energy linearity in temporal-distance}
\sigma _q^2 = \frac{1}{{{F_{rate}}}} \cdot \tilde \sigma _q^2
\end{IEEEeqnarray}
Where $\tilde \sigma _q^2$ is the energy of successive frames difference for a sequence of one frame per second.

\subsection{Compression Effect}
We model quality reduction due to compression as a component of the noise element $w_t$. This component is denoted as $w_{t,compression}$ and it is independent with other ingredients of $w_t$, which their sum is denoted as $w_{t,basic}$. As a result,
\begin{IEEEeqnarray}{rCl}
\label{eq:Compression effect abbreviated in w_t variance}
\sigma ^2 _w = \sigma ^2 _{w, compression} + \sigma ^2 _{w, basic}
\end{IEEEeqnarray}
where $\sigma ^2 _{w, compression}$ is the variance of the compression error, i.e., the mean-squared-error (MSE). $\sigma ^2 _{w, basic}$ is $w_{t,basic}$'s variance. 

We can express $\sigma ^2 _{w, compression}$ in various ways:
\subsubsection{Empirical rate-distortion curve}
\begin{IEEEeqnarray}{rCl}
\label{eq:Compression effect - w-compression variance - as rate distortion function}
\sigma ^2 _{w, compression} = \beta \cdot r ^ {-\alpha}
\end{IEEEeqnarray}
where, $\alpha$ and $\beta$ are curve parameters, and $r$ is the bit-rate.
\subsubsection{Theoretical rate-distortion estimation for memoryless Gaussian source}
A simple theoretical estimation is available under the following assumptions. Firstly, The compression distortion is similar to the procedure of directly compressing the frame pixels. Secondly, the frame pixels originated at a memoryless Gaussian source. The estimation is given by
\begin{IEEEeqnarray}{rCl}
\label{eq:Compression effect - w-compression variance - theoretical estimation}
\sigma ^2 _{w, compression} = \sigma _v^2\cdot{2^{ - 2r}},
\end{IEEEeqnarray}
where, $\sigma _v^2$ is the variance of the Gaussian source, and $r$ is the bit-rate.
\subsubsection{Given a value that is externally known or estimated}
\begin{IEEEeqnarray}{rCl}
\label{eq:Compression effect - w-compression variance - as compression MSE}
\sigma ^2 _{w, compression} = MSE_{compression}.
\end{IEEEeqnarray}
\subsubsection{Uncompressed video}
An uncompressed video has no compression error, i.e., $\sigma ^2 _{w, compression} = 0$; hence, $\sigma ^2 _{w} = \sigma ^2 _{w, basic}$.

\section{Analysis of Motion-Compensated Prediction}
\label{sec:Analysis of Motion-Compensated Prediction}
In this section we analyze the two common cases of applying motion-compensation. First, we consider MC-prediction between a pair of available frames, as in MC-coding. Then, we study the case of applying MC-prediction between an existing and absent frames, as in MC-FRUC.

Our analysis is statistical, therefore we can differ from practical MC-prediction as follows. First, we treat a single MC procedure for a given signal properties, since it is statistically representative. Second, we assume it is statistically allowed to consider signals as functions without explicit spatial boundaries, although a practical MC-prediction has defined dimensions for block and search areas.

\subsection{MC-Prediction of an Available Frame}
\label{sec:MC-Prediction of an Available Frame}
Let us consider the MC-prediction of frame $f_t$ using $f_{t-i}$ as a reference frame, where $i \in \left\{ {0,...,t - 1} \right\}$. The prediction relies on estimating the motion between $t-i$ and $t$ using the corresponding frames. This estimation assumes translational motion and is denoted as $\hat \varphi \left( {t,t - i\left| {{f_t},{f_{t - i}}} \right.} \right) = \left( {{{\hat \varphi }_x}\left( {t,t - i\left| {{f_t},{f_{t - i}}} \right.} \right),{{\hat \varphi }_y}\left( {t,t - i\left| {{f_t},{f_{t - i}}} \right.} \right)} \right)$. We describe the MC-prediction as follows,
\begin{IEEEeqnarray}{rCl}
\label{eq:MC-prediction - initial expression - ME using reference frames}
&& {\hat f_t}\left( {x,y\left| {{{f_{t - i}^{ref}}},} \right.\hat \varphi \left( {t,t - i\left| {{f_t},{f_{t - i}^{ref}}} \right.} \right)} \right) = 
\\ \nonumber
&& \qquad {} {f_{t - i}^{ref}}\left( {x - {{\hat \varphi }_x}\left( {t,t - i\left| {{f_t},{f_{t - i}^{ref}}} \right.} \right),y - {{\hat \varphi }_y}\left( {t,t - i\left| {{f_t},{f_{t - i}}} \right.} \right)} \right)
\end{IEEEeqnarray}
Where $f_{t - i}^{ref}$ is a processed or distorted version of $f_{t - i}$ that serves as a reference frame. A reference frame at time $t$ is defined as
\begin{equation}
\label{eq:frame decomposition equation - reference frame}
{f_t^{ref}}\left( {x,y} \right) = {v_t}\left( {x,y} \right) + {n_t^{ref}}\left( {x,y} \right) .
\end{equation}
Where, according to (\ref{eq:accumulated noise definition}), ${n_t^{ref}}$ contains $w_t^{ref}$ that expresses the reference frame distortions.
For example, real hybrid encoders utilize closed-loop MC-coding by using the reconstructed-from-compression version of $f_{t - i}$; hence, we can express $w_t^{ref}$'s variance using (\ref{eq:Compression effect abbreviated in w_t variance}).

We assume that
\begin{IEEEeqnarray}{rCl}
\label{eq:Compression does not affect ME assumption}
\hat \varphi \left( {t,t - i\left| {{f_t},{f_{t - i}^{ref}}} \right.} \right) \approx \hat \varphi \left( {t,t - i\left| {{f_t},{f_{t - i}}} \right.} \right),
\end{IEEEeqnarray}
i.e., compression does not affect ME accuracy significantly. Hence, (\ref{eq:MC-prediction - initial expression - ME using reference frames}) is modified to 
\begin{IEEEeqnarray}{rCl}
\label{eq:MC-prediction - initial expression}
&& {\hat f_t}\left( {x,y\left| {{{f_{t - i}^{ref}}},} \right.\hat \varphi \left( {t,t - i\left| {{f_t},{f_{t - i}}} \right.} \right)} \right) = 
\\ \nonumber
&& \qquad {} {f_{t - i}^{ref}}\left( {x - {{\hat \varphi }_x}\left( {t,t - i\left| {{f_t},{f_{t - i}}} \right.} \right),y - {{\hat \varphi }_y}\left( {t,t - i\left| {{f_t},{f_{t - i}}} \right.} \right)} \right).
\end{IEEEeqnarray}
The ME is approximated using (\ref{eq:Compression does not affect ME assumption}); however, the compression still affects the MC residual through $\sigma ^2 _{w,compression}$ of the reference frame.

We assume that the object from $f_t$, which its motion is estimated, is contained in the search area in $f_{t-i}$. Therefore, we model $\hat \varphi\left( {t,t-i\left| {{f_t},{f_{t - i}}} \right.} \right)$ to have a displacement error $\left( {\Delta x,\Delta y} \right)$ that depends only on the spatial properties of the ME algorithm, e.g., search resolution. Hence, the error excludes any temporal dependency. Specifically, 
\begin{IEEEeqnarray}{rCl}
\label{eq:MC-prediction - available frame motion estimation model}
{{\hat \varphi }_x}\left( {t,t - i\left| {{f_t},{f_{t - i}}} \right.} \right) & = & {\varphi _x}\left( {t,t - i} \right) + \Delta x
\\ \nonumber
{{\hat \varphi }_y}\left( {t,t - i\left| {{f_t},{f_{t - i}}} \right.} \right) & = & {\varphi _y}\left( {t,t - i} \right) + \Delta y
\end{IEEEeqnarray}
Where $\Delta x$ and $\Delta y$ are uniformly distributed in a range defined by the accuracy of the ME algorithm. 
Using (\ref{eq:frame decomposition equation}), (\ref{eq:v_t definition}) and (\ref{eq:MC-prediction - available frame motion estimation model}) we develop (\ref{eq:MC-prediction - initial expression}) into
\begin{IEEEeqnarray}{rCl}
\label{eq:MC-prediction - frame prediction model}
\nonumber && {{\hat f}_t}\left( {x,y\left| {{f_{t - i}^{ref}},} \right.\hat \varphi \left( {t,t - i\left| {{f_t},{f_{t - i}}} \right.} \right)} \right) = v\left( {x - {\varphi _x}\left( {t,0} \right) - \Delta x,y - {\varphi _y}\left( {t,0} \right) - \Delta y} \right)
\\ 
&& \qquad {} + {n_{t - i}^{ref}}\left( {x - {{\hat \varphi }_x}\left( {t,t - i\left| {{f_t},{f_{t - i}}} \right.} \right),y - {{\hat \varphi }_y}\left( {t,t - i\left| {{f_t},{f_{t - i}}} \right.} \right)} \right) .
\end{IEEEeqnarray}
Here we used the property $\varphi \left( {t,0} \right) = \varphi \left( {t,t - i} \right) + \varphi \left( {t - i,0} \right)$ that follows from the definition in (\ref{eq:motion between two frames}).

The MC-prediction error of $f_t$ using $f_{t-i}$ as a reference frame is formulated as
\begin{equation}
\label{eq:MC-prediction error}
{e_{t|t - i}}\left( {x,y} \right) = {f_t}\left( {x,y} \right) - {{\hat f}_t}\left( {x,y\left| {{f_{t - i}^{ref}},\hat \varphi \left( {t,t - i\left| {{f_t},{f_{t - i}}} \right.} \right)} \right.} \right)
\end{equation}
In appendix \ref{appendix_sec:MC Prediction of an Available Frame}, we describe in detail the calculation of the autocorrelation function of the MC-prediction error. This derivation results in 
\begin{IEEEeqnarray}{rCl}
\label{eq:Available frame error autocorrelation - implicit}
{R_{{e_i}}}\left( {k,l} \right) & = & 2\left( {\sigma _{\Delta x}^2 + \sigma _{\Delta y}^2} \right) \cdot \left[ {{R_v}\left( {k,l} \right) + {R_{{n_{t - i}^{ref}}}}\left( {k,l} \right)} \right]
\\ \nonumber
&& - \sigma _{\Delta x}^2\cdot\left[ {{R_v}\left( {k - 1,l} \right) + {R_v}\left( {k + 1,l} \right)} + {{R_{{n_{t - i}^{ref}}}}\left( {k - 1,l} \right) + {R_{{n_{t - i}^{ref}}}}\left( {k + 1,l} \right)} \right]
\\ \nonumber
&& - \sigma _{\Delta y}^2\cdot\left[ {{R_v}\left( {k,l - 1} \right) + {R_v}\left( {k,l + 1} \right)} +  {{R_{{n_{t - i}^{ref}}}}\left( {k,l - 1} \right) + {R_{{n_{t - i}^{ref}}}}\left( {k,l + 1} \right)} \right]
\\ \nonumber
&& + {R_{\Delta {n_{t,t - i}}}}\left( {k,l} \right)
\end{IEEEeqnarray}
where ${R_{\Delta {n_{t,t - i}}}}\left( {k,l} \right)$ is the autocorrelation of the MC noise difference, denoted as $\Delta {n_{t_2,t_1}}$ for $t_1 < t_2$ and defines as
\begin{IEEEeqnarray}{rCl}
\label{eq:MC-prediction - motion-compensated noise difference - definition}
&& \Delta {n_{t_2,t_1}}\left( {x,y} \right) \equiv {n_{t_2}}\left( {x,y} \right) - {n_{t_1}^{ref}}\left( {x - {\varphi _x}\left( {t_2,t_1} \right),y - {\varphi _y}\left( {t_2,t_1} \right)} \right)
\end{IEEEeqnarray}

The following autocorrelation was calculated for $\Delta {n_{t_2,t_1}}$ in the appendix (\ref{eq:MC-prediction - motion-compensated noise difference - calculation})-(\ref{eq:MC-prediction - motion-compensated noise difference - autocorrelation}):
\begin{IEEEeqnarray}{rCl}
\label{eq:MC-prediction - motion-compensated noise difference - autocorrelation - short}
\nonumber {R_{\Delta {n_{{t_2},{t_1}}}}}\left( {k,l} \right) & = & \left[ {2\sigma _q^2\cdot\left( {{t_2} - {t_1}} \right) + \sigma _{w _{t_1}}^2 + \sigma _{w _{t_2}}^2 } \right] \cdot \delta \left( {k,l} \right)
\\
\end{IEEEeqnarray}

The following explicit form of (\ref{eq:Available frame error autocorrelation - implicit}) is provided in the appendix:
\begin{IEEEeqnarray}{rCl}
\label{eq:MC-prediction - available frame error autocorrelation - explicit}
{R_{{e_i}}}\left( {k,l} \right) & = & 2\left[ {\sigma _{\Delta x}^2 + \sigma _{\Delta y}^2} \right]\cdot\left[ {\sigma _v^2\cdot\rho _v^{\left| k \right| + \left| l \right|} + \left( {L \sigma _q^2 + \sigma _{w,ref}^2} \right)\cdot\delta \left( {k,l} \right)} \right]
\\ \nonumber
&& - \sigma _{\Delta x}^2\sigma _v^2{\rho_v ^{\left| l \right|}} \cdot \left[ {{\rho_v ^{\left| {k - 1} \right|}} + {\rho_v^{\left| {k + 1} \right|}}} \right]
\\ \nonumber
&& - \sigma _{\Delta x}^2\left[ {L\sigma _q^2 + \sigma _{w,ref}^2} \right]\cdot\left[ {\delta \left( {k - 1,l} \right) + \delta \left( {k + 1,l} \right)} \right]
\\ \nonumber
&& - \sigma _{\Delta y}^2\sigma _v^2{\rho_v^{\left| k \right|}} \cdot \left[ {{\rho_v ^{\left| {l - 1} \right|}} + {\rho_v^{\left| {l + 1} \right|}}} \right]
\\ \nonumber
&& - \sigma _{\Delta y}^2\left[ {L\sigma _q^2 + \sigma _{w,ref}^2} \right]\cdot\left[ {\delta \left( {k,l - 1} \right) + \delta \left( {k,l + 1} \right)} \right]
\\ \nonumber
&& + \left[ {2i\sigma _q^2 + \sigma _{w,current}^2 + \sigma _{w,ref}^2} \right]\cdot\delta \left( {k,l} \right)
\end{IEEEeqnarray}
The error variance is
\begin{IEEEeqnarray}{rCl}
\label{eq:MC-prediction - available frame error variance}
{R_{{e_i}}}\left( {0,0} \right) & = & 2\left( {\sigma _{\Delta x}^2 + \sigma _{\Delta y}^2} \right)\cdot\left[ {\sigma _v^2\cdot\left( {1 - {\rho _v}} \right) + \left( {L\sigma _q^2 + \sigma _{w,ref}^2} \right)} \right]
\\ \nonumber
&& + {2i\sigma _q^2 + \sigma _{w,current}^2 + \sigma _{w,ref}^2}
\end{IEEEeqnarray}
The last expression shows a linear relation between the variance and the temporal-distance represented here in frame units, $i$. Translation of the temporal-distance to seconds (denoted as $d_t$) is possible using (\ref{eq:Difference energy linearity in temporal-distance}):
\begin{IEEEeqnarray}{rCl}
\label{eq:MC-prediction - available frame error variance - temporal distance in seconds}
{R_{{e_i}}}\left( {0,0} \right) & = & 2\left( {\sigma _{\Delta x}^2 + \sigma _{\Delta y}^2} \right)\cdot\left[ {\sigma _v^2\cdot\left( {1 - {\rho _v}} \right) + \left( {\frac{L}{{{F_{rate}}}}\tilde \sigma _q^2 + \sigma _{w,ref}^2} \right)} \right]
\\ \nonumber
&& + 2\tilde \sigma _q^2{d_t} + \sigma _{w,current}^2 + \sigma _{w,ref}^2
\end{IEEEeqnarray}

\subsection{MC-Prediction of an Absent Frame}
\label{subsec:MC-Prediction of an Absent Frame}
Let us consider temporal upsampling by a factor of $D$ using MC-FRUC, i.e., $D-1$ missing frames are interpolated between each two existing frames. The available frames are denoted as $f_0$ and $f_D$, and the interpolated frames are denoted as $\left\{ {\hat f_j^{}} \right\}_{j = 1}^{D - 1}$. We consider the interpolation of a block in the $j^{th}$ interpolated frame, where $j \in \left\{ {1,...,D - 1} \right\}$.
The corresponding unavailable frame is denoted as $f_j$.

The prediction includes estimation of the motion between the $j^{th}$ frame and each of the available frames, $f_0$ and $f_D$. The estimation is performed using $f_0$ and $f_D$. $\hat \varphi \left( {j,0\left| {{f_0},{f_D}} \right.} \right)$ and $\hat \varphi \left( {D,j\left| {{f_0},{f_D}} \right.} \right)$ denote the estimated motion at $f_j$ relative to frames $f_0$ and $f_D$, respectively. We assume
\begin{IEEEeqnarray}{rCl}
\label{eq:MC-prediction - absent frame - motion prediction}
&& \hat \varphi \left( {j,0\left| {{f_0},{f_D}} \right.} \right) = \left( {{\varphi _x}\left( {j,0} \right) + \Delta x_0^{abs},{\varphi _y}\left( {j,0} \right) + \Delta y_0^{abs}} \right)
\\ \nonumber
&& \hat \varphi \left( {D,j\left| {{f_0},{f_D}} \right.} \right) = \left( {{\varphi _x}\left( {D,j} \right) + \Delta x_D^{abs},{\varphi _y}\left( {D,j} \right) + \Delta y_D^{abs}} \right)
\end{IEEEeqnarray}
Where $\Delta x_0^{abs}$ and $\Delta x_D^{abs}$ are assumed to be independent Gaussian random variables with zero-mean and variance
\begin{IEEEeqnarray}{rCl}
\label{eq:MC-prediction - absent frame - motion prediction spatial error assumption}
\sigma _{\Delta {x^{abs}}}^2 = \gamma _{abs} \cdot \sigma _{\Delta x}^2.
\end{IEEEeqnarray}
Where $\sigma _{\Delta x}^2$ is the variance of $\Delta x$, which was defined above for the case of an available frame, and $\gamma _{abs} > 1$ denotes effect of the frame absence on the spatial accuracy of the ME. $\Delta y_0^{abs}$ and $\Delta y_D^{abs}$ are defined accordingly by replacing $x$ with $y$.

The overall prediction is calculated using two prediction signals. The backward prediction is defined as
\begin{IEEEeqnarray}{rCl}
\label{eq:MC-prediction - absent frame - f0 prediction}
\nonumber && {{\hat f}_j}\left( {x,y\left| {{f_0},} \right.\hat \varphi \left( {j,0\left| {{f_0},{f_D}} \right.} \right)} \right) = {f_0}\left( {x - {{\hat \varphi }_x}\left( {j,0\left| {{f_0},{f_D}} \right.} \right),y - {{\hat \varphi }_y}\left( {j,0\left| {{f_0},{f_D}} \right.} \right)} \right).
\\
\end{IEEEeqnarray}
and the forward prediction as
\begin{IEEEeqnarray}{rCl}
\label{eq:MC-prediction - absent frame - fD prediction}
\nonumber && {{\hat f}_j}\left( {x,y\left| {{f_D},} \right.\hat \varphi \left( {D,j\left| {{f_0},{f_D}} \right.} \right)} \right) = {f_D}\left( {x + {{\hat \varphi }_x}\left( {D,j\left| {{f_0},{f_D}} \right.} \right),y + {{\hat \varphi }_y}\left( {D,j\left| {{f_0},{f_D}} \right.} \right)} \right).
\\
\end{IEEEeqnarray}

The final prediction is achieved by the following linear combination of (\ref{eq:MC-prediction - absent frame - f0 prediction}) and (\ref{eq:MC-prediction - absent frame - fD prediction}):
\begin{IEEEeqnarray}{rCl}
\label{eq:MC-prediction - absent frame - final prediction}
\hat f_j^{final}\left( {x,y\left| {{f_0},} \right.{f_D}} \right) & = & \theta  \cdot {{\hat f}_j}\left( {x,y\left| {{f_0},} \right.\hat \varphi \left( {j,0\left| {{f_0},{f_D}} \right.} \right)} \right)
\nonumber \\
&& + \left[ {1 - \theta } \right] \cdot {{\hat f}_j}\left( {x,y\left| {{f_D},} \right.\hat \varphi \left( {D,j\left| {{f_0},{f_D}} \right.} \right)} \right)
\end{IEEEeqnarray}
and the prediction error is expressed as 
\begin{IEEEeqnarray}{rCl}
\label{eq:MC-prediction - absent frame error expression}
&& e_{j|0,D}^{absent}\left( {x,y} \right) = {f_j}\left( {x,y} \right) - \hat f_j^{final}\left( {x,y\left| {{f_0},} \right.{f_D}} \right)
\end{IEEEeqnarray}

In appendix \ref{appendix_sec:MC Prediction of an Absent Frame}, we describe in detail the calculation of the autocorrelation function of the MC-prediction error. This derivation results in
\begin{IEEEeqnarray}{rCl}
\label{eq:Absent frame error autocorrelation - simplified form}
{R_{e_{j|0,D}^{absent}}}\left( {k,l} \right) & = & {\theta ^2} \cdot {R_{\Delta {n_{j,0}}}}\left( {k,l} \right) + {\left( {1 - \theta } \right)^2} \cdot {R_{\Delta {n_{D,j}}}}\left( {k,l} \right) + 
\\ \nonumber
&& + \sigma _{\Delta {x^{abs}}}^2 \cdot \left[ {{\theta ^2} + {{\left( {1 - \theta } \right)}^2}} \right] 
\\ \nonumber
&& \qquad \times \left[ {2{R_v}\left( {k,l} \right) - {R_v}\left( {k - 1,l} \right) - {R_v}\left( {k + 1,l} \right)} \right.
\\ \nonumber
&& \qquad\qquad {} \left. + {2{R_{{n_0}}}\left( {k,l} \right) - {R_{{n_0}}}\left( {k - 1,l} \right) - {R_{{n_0}}}\left( {k + 1,l} \right)} \right]
\\ \nonumber
&& + \sigma _{\Delta {y^{abs}}}^2 \cdot \left[ {{\theta ^2} + {{\left( {1 - \theta } \right)}^2}} \right] 
\\ \nonumber
&& \qquad \times \left[ {2{R_v}\left( {k,l} \right) - {R_v}\left( {k,l - 1} \right) - {R_v}\left( {k,l + 1} \right)} \right.
\\ \nonumber
&& \qquad\qquad {} \left. + {2{R_{{n_0}}}\left( {k,l} \right) - {R_{{n_0}}}\left( {k,l - 1} \right) - {R_{{n_0}}}\left( {k,l + 1} \right)} \right]
\end{IEEEeqnarray}

Let us study the variance of the error. This variance is also the mean-squared error (MSE) of the interpolation procedure; hence, it is useful for performance evaluation in applications such as FRUC. 
Using (\ref{eq:pure image autocorrelation}),(\ref{eq:accumulated noise autocorrelation}) and (\ref{eq:MC-prediction - motion-compensated noise difference - autocorrelation - short}),
we calculate from (\ref{eq:Absent frame error autocorrelation - simplified form}) the following MSE expression.
\begin{IEEEeqnarray}{rCl}
\label{eq:MC-prediction - absent frame - error variance - non static region}
{R_{e_{j|0,D}^{absent}}}\left( {0,0} \right) & = & {\theta ^2}\cdot\left[ {2\sigma _q^2j + \sigma _{{w_0}}^2 + \sigma _{{w_j}}^2} \right] \\ \nonumber
&& + {\left( {1 - \theta } \right)^2}\cdot\left[ {2\sigma _q^2\left( {D - j} \right) + \sigma _{{w_0}}^2 + \sigma _{{w_j}}^2} \right]
\\ \nonumber
&& + 2\left( {\sigma _{\Delta {x^{abs}}}^2 + \sigma _{\Delta {y^{abs}}}^2} \right)\cdot\left[ {{\theta ^2} + {{\left( {1 - \theta } \right)}^2}} \right] \cdot \left[ {\left( {1 - {\rho _v}} \right)\cdot\sigma _v^2 + L\sigma _q^2 + \sigma _{{w_0}}^2} \right]
\end{IEEEeqnarray}
Usually, $\theta$ is set to $0.5$ for the central part of the interpolated block. We assume $\theta = 0.5$ for the entire interpolated area; hence, (\ref{eq:MC-prediction - absent frame - error variance - non static region}) becomes
\begin{IEEEeqnarray}{rCl}
\label{eq:MC-prediction - absent frame - error variance - non static region - for theta 0.5}
{R_{e_{j|0,D}^{absent}}}\left( {0,0} \right) & = & \frac{1}{2}\cdot\left[ {\sigma _q^2D + \sigma _{{w_0}}^2 + \sigma _{{w_j}}^2} \right] 
\\ \nonumber
&& + \left( {\sigma _{\Delta {x^{abs}}}^2 + \sigma _{\Delta {y^{abs}}}^2} \right)\cdot\left[ {\left( {1 - {\rho _v}} \right)\cdot\sigma _v^2 + L\sigma _q^2 + \sigma _{{w_0}}^2} \right] .
\end{IEEEeqnarray}
The last expression shows that the variance is a linear function of the temporal-distance between the available frames, $D$. Moreover, according to (\ref{eq:Difference energy linearity in temporal-distance}), the linear relation with $\sigma _q^2$ implies a linear relation with the basic temporal-distance derived from the frame-rate. In addition, Recall that $\sigma _{\Delta {x^{abs}}}^2$ and $\sigma _{\Delta {y^{abs}}}^2$ are linear functions of $D$ (\ref{eq:MC-prediction - absent frame - motion prediction spatial error assumption}).

FRUC application may be applied on processed or reconstructed-from-compression video. The quality of the video affects FRUC performance. Our model supports these cases through the noise component of $f_0$ and $f_D$ frames; i.e., by including the processed video's MSE in $\sigma _{{w_0}}^2$ and $\sigma _{{w_D}}^2$, as in (\ref{eq:Compression effect abbreviated in w_t variance}) and (\ref{eq:Compression effect - w-compression variance - as compression MSE}).

\section{A Simplified Autocorrelation Model for MC-Prediction Error in Coding }
\label{sec:Simplified Autocorrelation Model}
In section \ref{sec:MC-Prediction of an Available Frame}, we proposed an autocorrelation function for the error of MC-prediction of an available frame, as in coding applications. The proposed autocorrelation function (\ref{eq:MC-prediction - available frame error autocorrelation - explicit}) is rather complicated.
Therefore, it may be useful to have also a simpler autocorrelation model.
In this section, we propose a simpler autocorrelation model for the MC-residual in coding systems. The autocorrelation of MC-FRUC can be simplified similarly; however, it is unnecessary since FRUC analysis usually considers only the variance, which is equal to the interpolation MSE.

\subsection{General Construction of A Separable Model}
Similarly to \cite{RefWorks:60} and \cite{RefWorks:10}, we construct a model of a separable form from the complicated autocorrelation function. As a result, the linearity of the variance in the temporal-distance is kept.

The variance-normalized autocorrelation function (ACF) is defined as
\begin{equation}
\label{eq:variance-normalized autocorrelation function}
{\rho _{{e_i}}}\left( {k,l} \right) = \frac{{{R_{{e_i}}}\left( {k,l} \right)}}{{{R_{{e_i}}}\left( {0,0} \right)}}.
\end{equation}
The variance-normalized ACF along the horizontal axis is defined as
\begin{equation}
\label{eq:variance-normalized autocorrelation function along horizontal axis}
\rho _{{e_i}}^{horz}\left( k \right) = \frac{{{R_{{e_i}}}\left( {k,0} \right)}}{{{R_{{e_i}}}\left( {0,0} \right)}}.
\end{equation}
The variance-normalized ACF along the vertical axis is defined correspondingly and denoted as $\rho _{{e_i}}^{vert}\left( l \right)$.

A separable form of ${{R_{{e_i}}}\left( {k,l} \right)}$ is formed as follows:
\begin{IEEEeqnarray}{rCl}
\label{eq:A separable form of an autocorrelation function}
R_{{e_i}}^{sep}\left( {k,l} \right) = {R_{{e_i}}}\left( {0,0} \right) \cdot \rho _{{e_i}}^{horz}\left( k \right)\rho _{{e_i}}^{vert}\left( l \right)
\end{IEEEeqnarray}

Let us derive a separable model in the form of (\ref{eq:A separable form of an autocorrelation function}) for the autocorrelation function given in (\ref{eq:MC-prediction - available frame error autocorrelation - explicit}). This requires the calculation of ${{R_{{e_i}}}\left( {0,0} \right)}$, $\rho _{{e_i}}^{horz}\left( k \right)$ and $\rho _{{e_i}}^{vert}\left( l \right)$ that correspond to (\ref{eq:MC-prediction - available frame error variance}). The variance ${{R_{{e_i}}}\left( {0,0} \right)}$ is given in (\ref{eq:MC-prediction - available frame error variance}).

\begin{figure*}[!t]
\centering
{\subfloat[]{\includegraphics[width=2.2in]{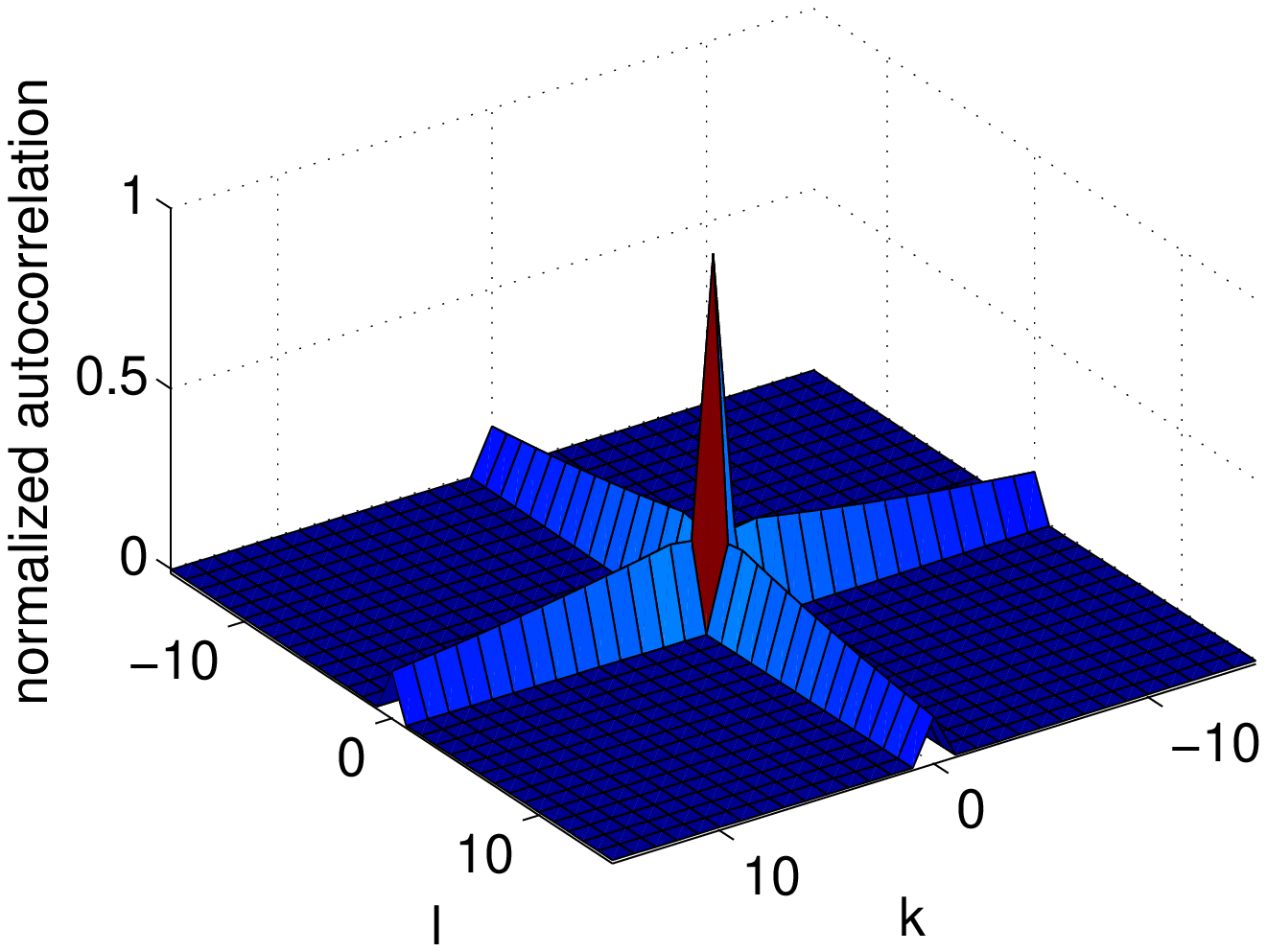}
\label{Fig:MC_coding_residual_autocorrelation_estimation_full}}}
\subfloat[]{\includegraphics[width=2.2in]{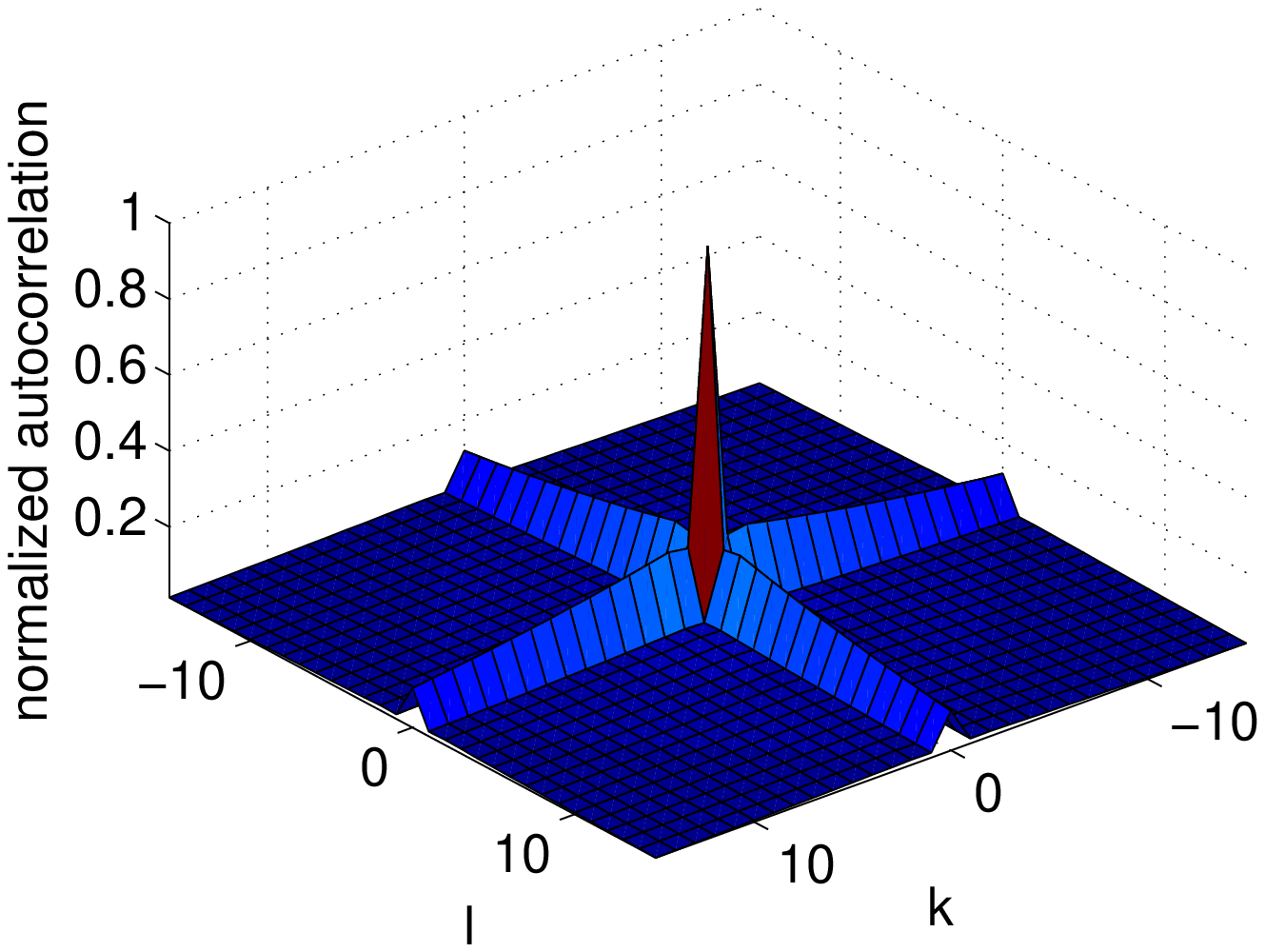}
\label{Fig:MC_coding_residual_autocorrelation_estimation_separable_construction}}
{\subfloat[]{\includegraphics[width=2.2in]{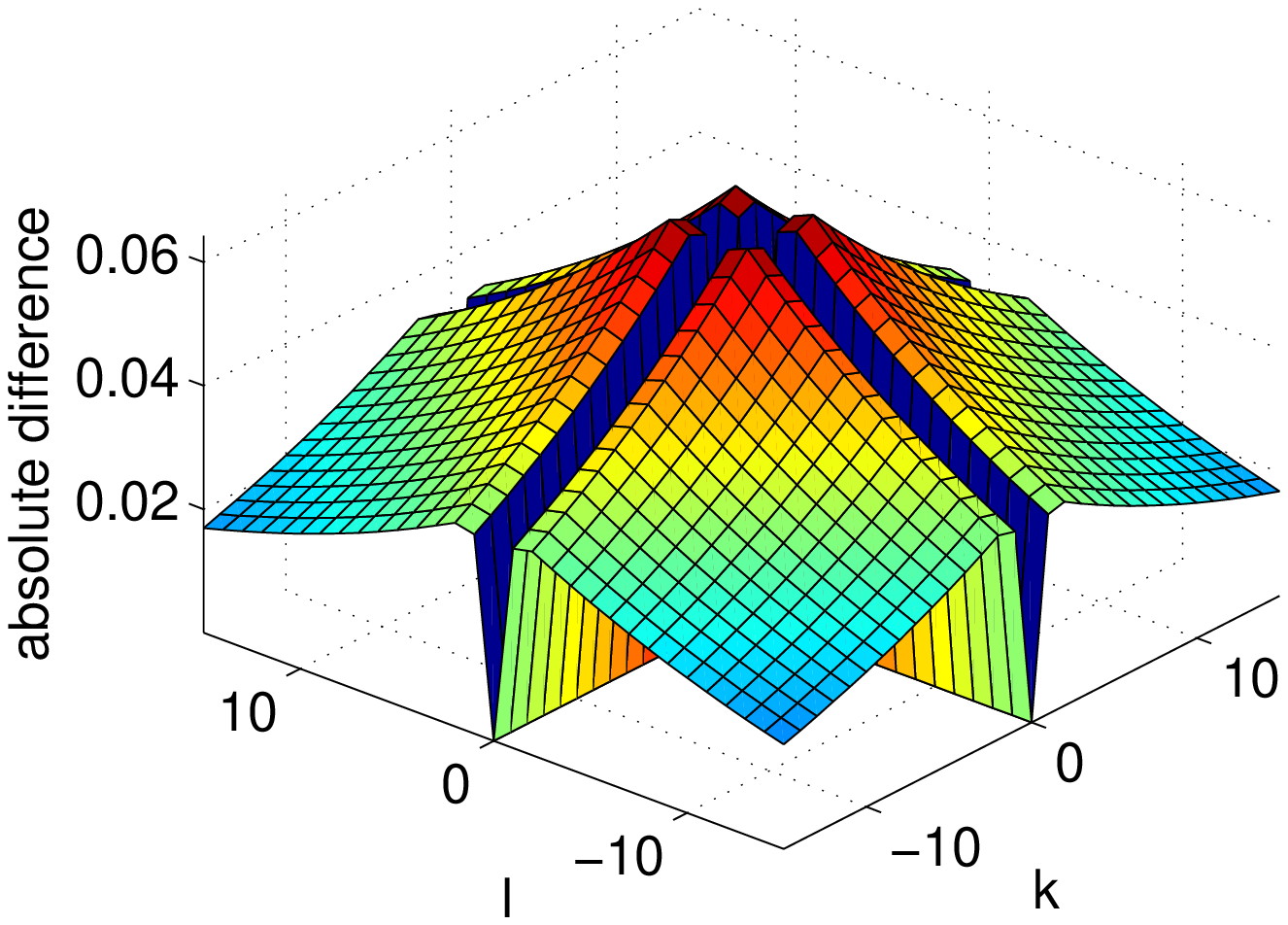}
\label{Fig:MC_coding_residual_autocorrelation_estimation_separable_construction__absolute_difference}}}
\caption{Estimation of MC-residual autocorrelation in MC-coding. (a) full model (\ref{eq:MC-prediction - available frame error autocorrelation - explicit}). (b) simplified model using separable construction (\ref{eq:A separable form of an autocorrelation function}). (c) absolute difference due to simplification.}
\label{Fig:Estimation of MC-residual autocorrelation in MC-coding}
\end{figure*}

First, we calculate ${R_{{e_i}}}\left( {k,0} \right)$ as follows.
\begin{IEEEeqnarray}{rCl}
\label{eq:Non-normalized horizontal autocorrelation for MC-prediction of an available frame}
{R_{{e_i}}}\left( {k,0} \right) & = & 2\left[ {\sigma _{\Delta x}^2 + \sigma _{\Delta y}^2} \right]\cdot\left[ {\sigma _v^2\cdot\rho _v^{\left| k \right|} + \left( {L \sigma _q^2 + \sigma _{w,ref}^2} \right) \cdot\delta \left( k \right)} \right]
\\ \nonumber
&& - \sigma _{\Delta x}^2\sigma _v^2\cdot\left[ {\rho _v^{\left| {k - 1} \right|} + \rho _v^{\left| {k + 1} \right|}} \right]
\\ \nonumber
&& - \sigma _{\Delta x}^2 \cdot \left( {L \sigma _q^2 + \sigma _{w,ref}^2} \right) \cdot\left[ {\delta \left( {k - 1} \right) + \delta \left( {k + 1} \right)} \right]
\\ \nonumber
&& - 2\sigma _{\Delta y}^2\sigma _v^2\rho _v^{\left| k \right| + 1} + \left( {2i\sigma _q^2 + \sigma _{w,current}^2 + \sigma _{w,ref}^2} \right) \cdot \delta \left( k \right)
\end{IEEEeqnarray}
Then, we get $\rho _{{e_i}}^{horz}\left( k \right)$ by dividing the last expression by ${{R_{{e_i}}}\left( {0,0} \right)}$ given in (\ref{eq:MC-prediction - available frame error variance}). $\rho _{{e_i}}^{vert}\left( l \right)$ is achieved similarly by replacing $x$ and $k$ with $y$ and $l$, respectively. Visual comparison of the original and simplified autocorrelation (Figs. \ref{Fig:MC_coding_residual_autocorrelation_estimation_full}, \ref{Fig:MC_coding_residual_autocorrelation_estimation_separable_construction}) shows high similarity while having acceptable differences (Fig. \ref{Fig:MC_coding_residual_autocorrelation_estimation_separable_construction__absolute_difference}).

\subsection{A Separable First-Order Markov Model}
While the autocorrelation function (\ref{eq:MC-prediction - available frame error autocorrelation - explicit}) was simplified to be separable (\ref{eq:A separable form of an autocorrelation function}), some users of the model may benefit from further simplification of the axis-ACF functions (e.g., (\ref{eq:Non-normalized horizontal autocorrelation for MC-prediction of an available frame})). We propose here to construct the autocorrelation function as a separable first-order Markov model. As a result, the horizontal and vertical autocorrelation functions will be exponential, i.e.,
\begin{IEEEeqnarray}{rCl}
\label{eq:First-order markov simplification of autocorrelation}
R_{{e_i}}^{Markov}\left( {k,l} \right) = {R_{{e_i}}}\left( {0,0} \right) \cdot \rho _{h,{e_i}} ^{\left| k \right|}\rho _{v,{e_i}}^{{\left| l \right|}}.
\end{IEEEeqnarray}
Where $R_{{e_i}}$ and ${R_{{e_i}}}\left( {0,0} \right)$ are the autocorrelation and variance of the accurate model (\ref{eq:MC-prediction - available frame error autocorrelation - explicit}). We define the correlation coefficients as follows,
\begin{IEEEeqnarray}{rCl}
\label{eq:First-order markov simplification of autocorrelation - correlation coefficients}
\rho _{h,{e_i}} = \frac{{{R_{{e_i}}}\left( {1,0} \right)}}{{{R_{{e_i}}}\left( {0,0} \right)}}\,\,\,\,\,\,\,\,\,\,\,\,and\,\,\,\,\,\,\,\,\,\,\,\, \rho _{v,{e_i}} = \frac{{{R_{{e_i}}}\left( {0,1} \right)}}{{{R_{{e_i}}}\left( {0,0} \right)}}.
\end{IEEEeqnarray}

\begin{figure*}[!t]
\centering
{\subfloat[]{\includegraphics[width=2.5in]{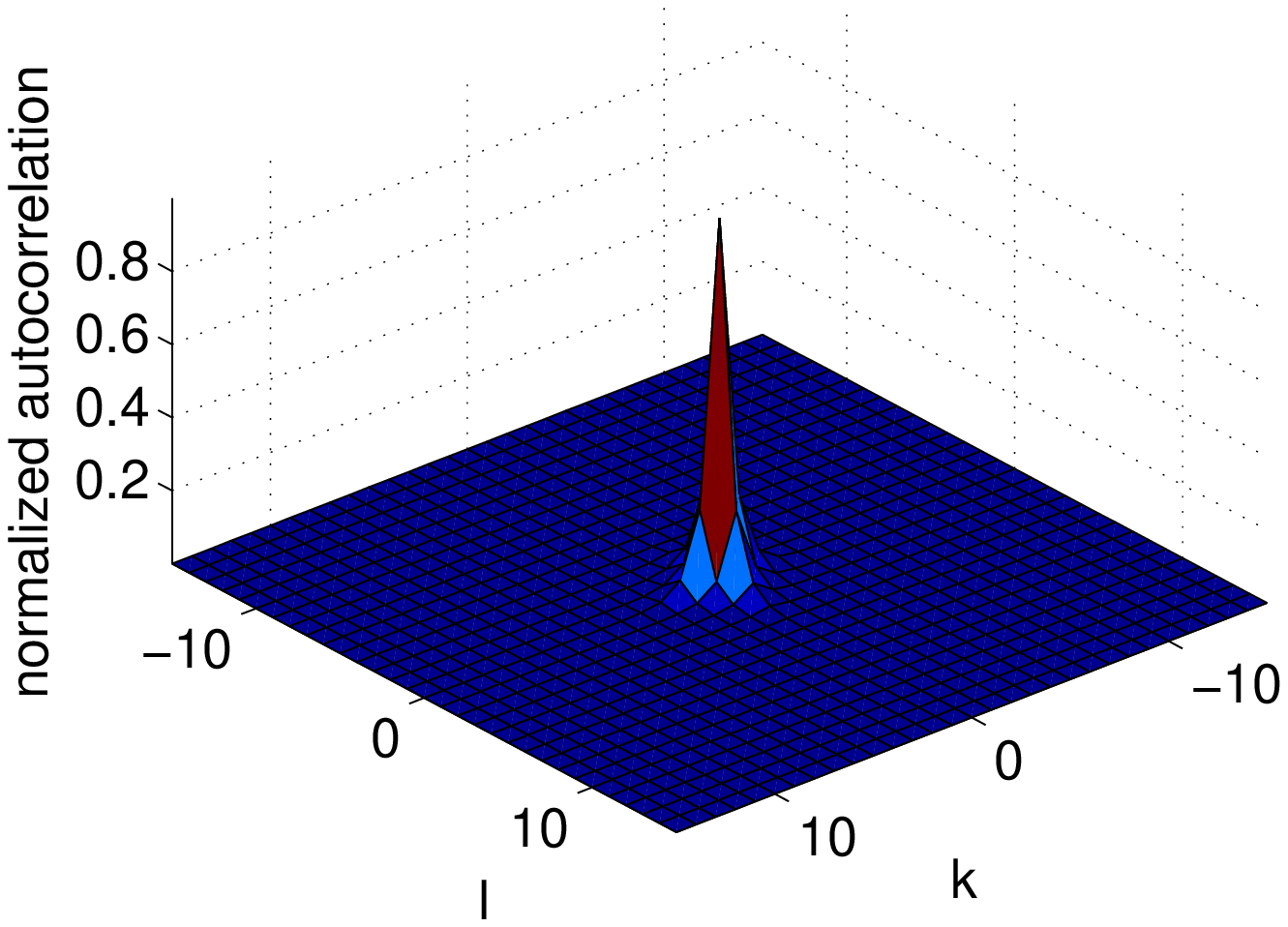}
\label{Fig:MC_coding_residual_autocorrelation_estimation_Markov}}}
\subfloat[]{\includegraphics[width=2.5in]{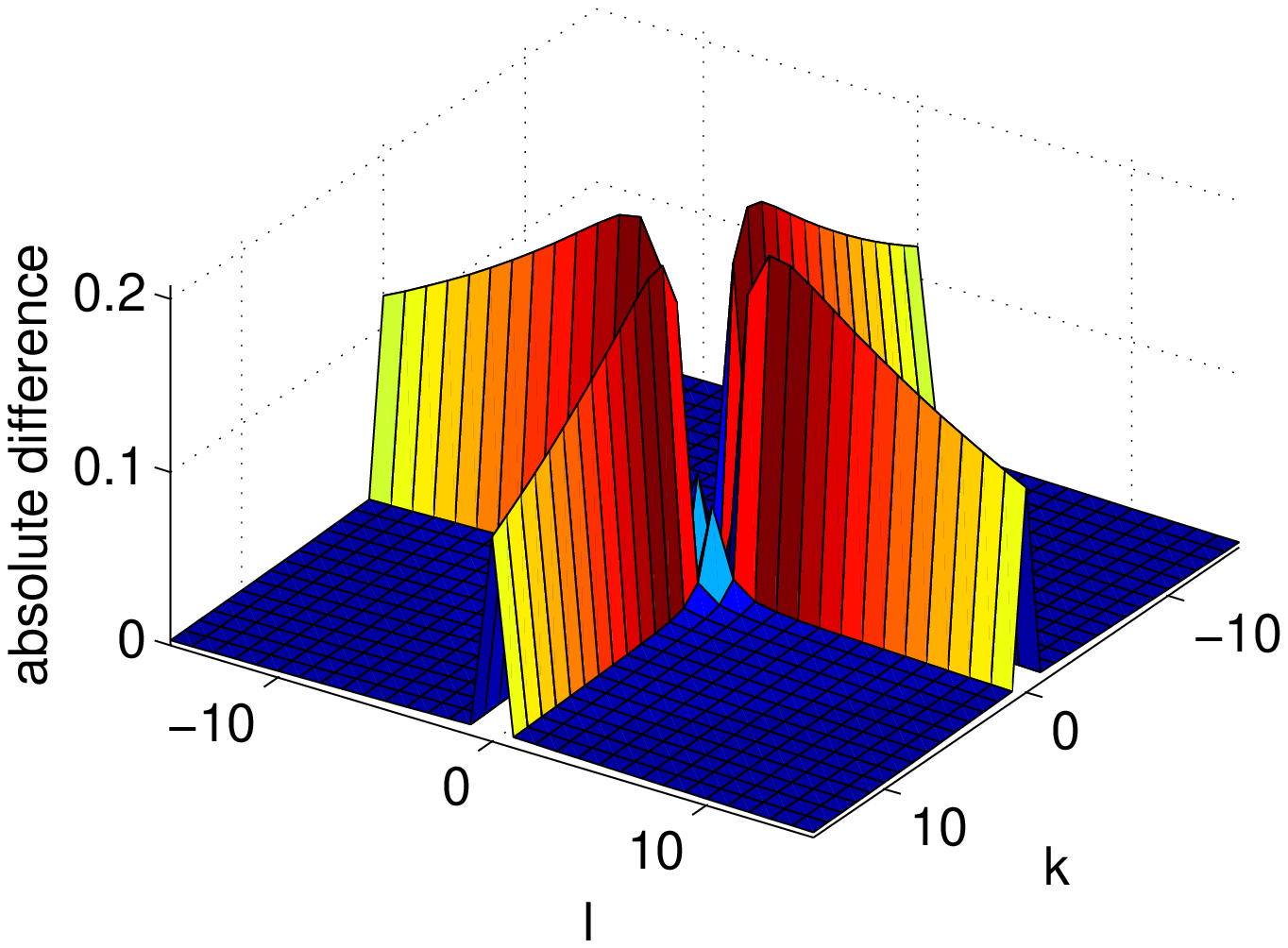}
\label{Fig:MC_coding_residual_autocorrelation_estimation_Markov__absolute_difference}}
\caption{Simplified autocorrelation function using first-order Markov model. (a) normalized autocorrelation. (b) absolute difference from full model (\ref{eq:MC-prediction - available frame error autocorrelation - explicit}).}
\label{Fig:Estimation of MC-residual autocorrelation in MC-coding - Markov simplification}
\end{figure*}

This model differs from the accurate model (\ref{eq:MC-prediction - available frame error autocorrelation - explicit}) and the previous simplification (\ref{eq:A separable form of an autocorrelation function}) in its lower values along the horizontal and vertical axes (Fig. \ref{Fig:Estimation of MC-residual autocorrelation in MC-coding - Markov simplification}). However, for coordinates that are not on the main axes, the difference from the accurate model is small (Fig. \ref{Fig:MC_coding_residual_autocorrelation_estimation_Markov__absolute_difference}), even more than in the former simplified model (Fig. \ref{Fig:MC_coding_residual_autocorrelation_estimation_separable_construction__absolute_difference}). In general, we consider this Markov model as an acceptable estimation when its added simplicity is needed.

\section{Theoretical Estimations}
\label{sec:Theoretical Estimations}
In this section we explore our model behavior for variation in main characteristics of the video signal and the compression procedure. We set $\sigma ^2 _v = 2312$, $\rho _v = 0.95$ and $L=5$.
The local noise component, $\sigma ^2 _w$, was calculated as follows. $\sigma ^2 _{w,basic}$ was set to zero, whereas $\sigma ^2 _{w,compression}$ was calculated according to (\ref{eq:Compression effect - w-compression variance - as rate distortion function}) with $\alpha = 1$ and $\beta = 10$.
We assume ME in half-pel accuracy; therefore, $\Delta x,\Delta y \in \left[ { - 0.25,0.25} \right]$ and  $\sigma _{\Delta x}^2 = \sigma _{\Delta y}^2 = {{{{\left( {2 \times 0.25} \right)}^2}} \mathord{\left/
 {\vphantom {{{{\left( {2 \times 0.25} \right)}^2}} {12}}} \right.
 \kern-\nulldelimiterspace} {12}}$.

\subsection{Motion-Compensated Coding}
\label{subsec:Theoretical Estimations - Motion-Compensated Coding}
\begin{figure*}[!t]
\centering
{\subfloat[]{\includegraphics[width=2.3in]{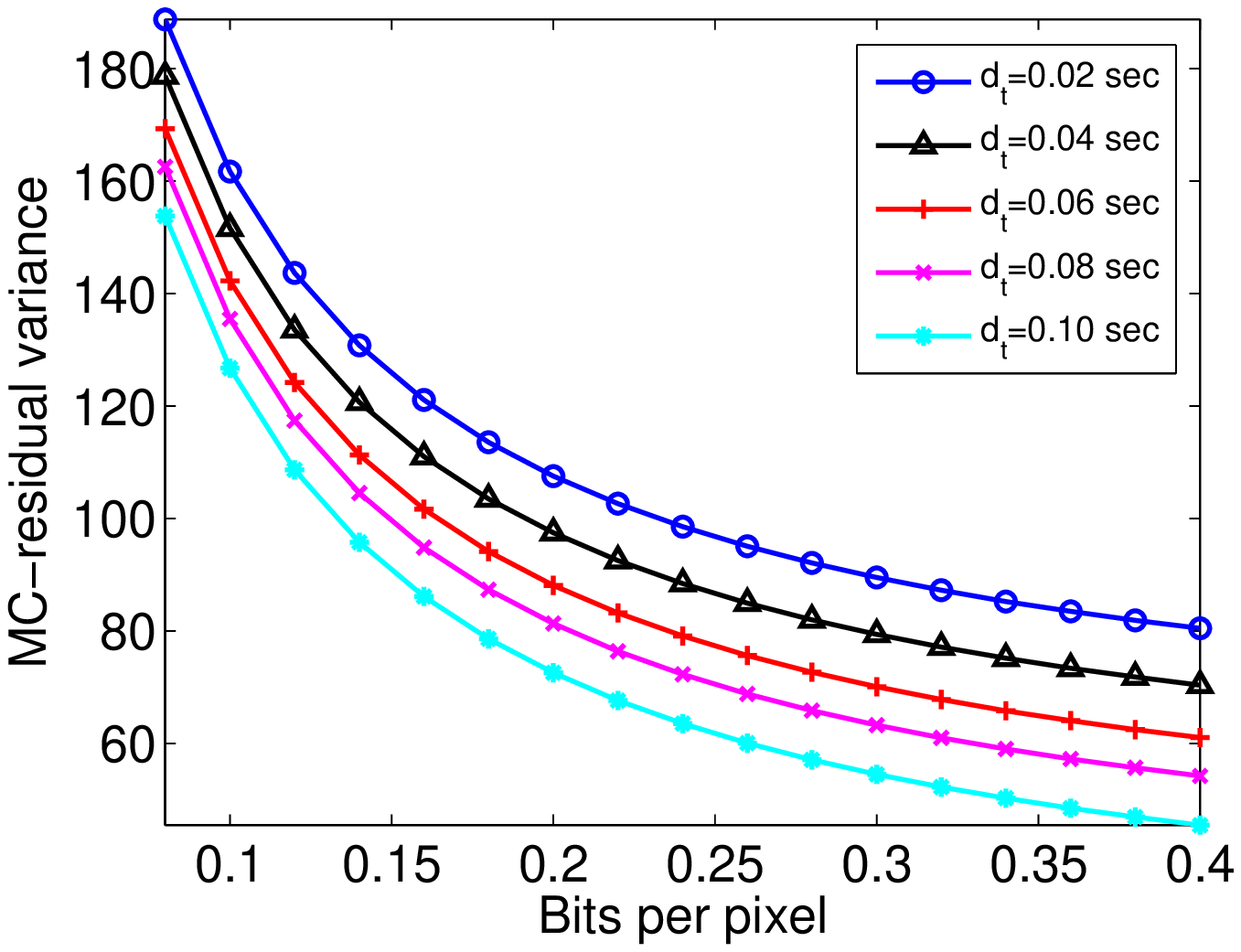}
\label{Fig:MC_coding_residual_variance_estimation_vs_bpp_for_various_framerates}}}
\subfloat[]{\includegraphics[width=2.3in]{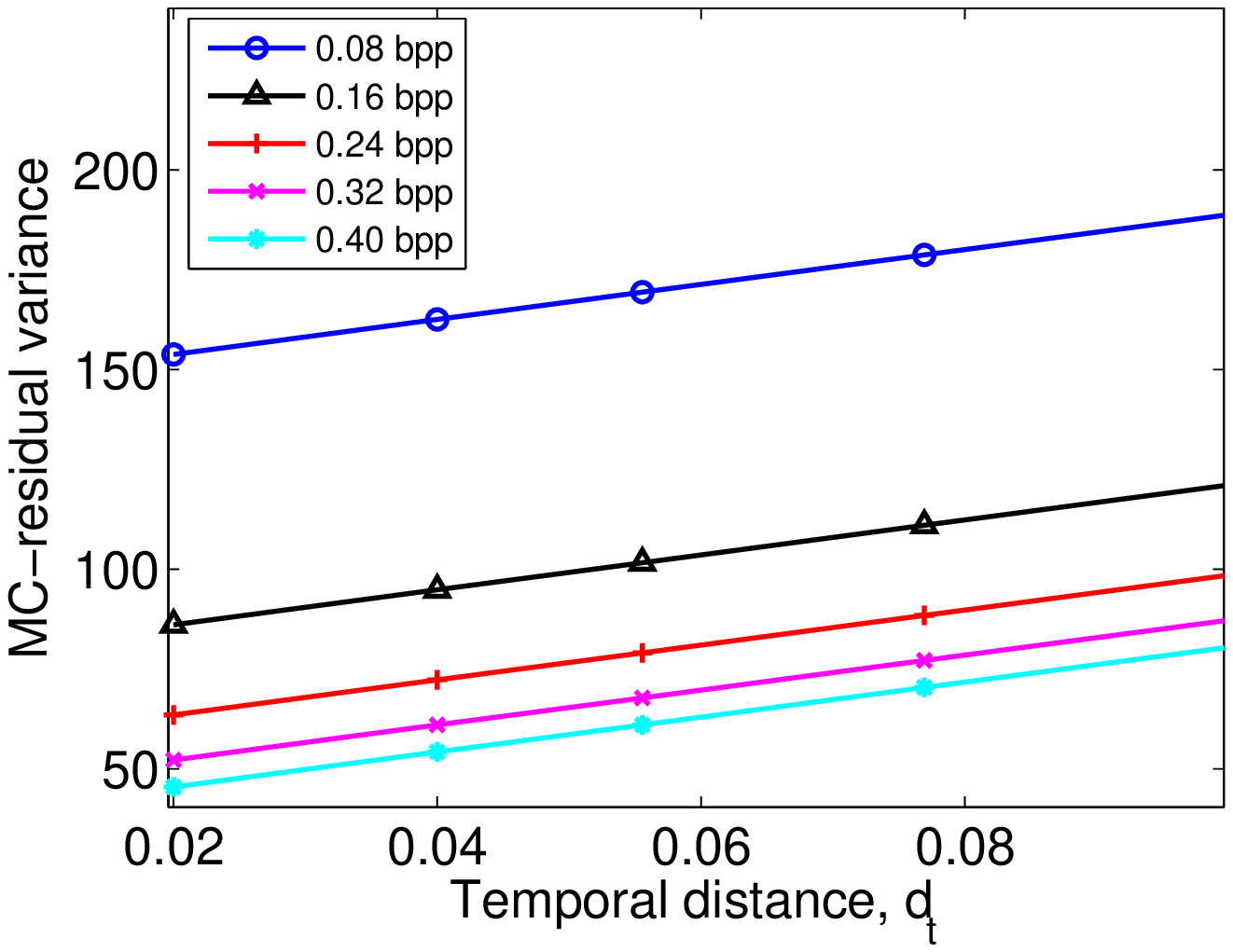}
\label{Fig:MC_coding_residual_variance_estimation_vs_temporal_distance_for_various_bpps}}
\subfloat[]{\includegraphics[width=2.3in]{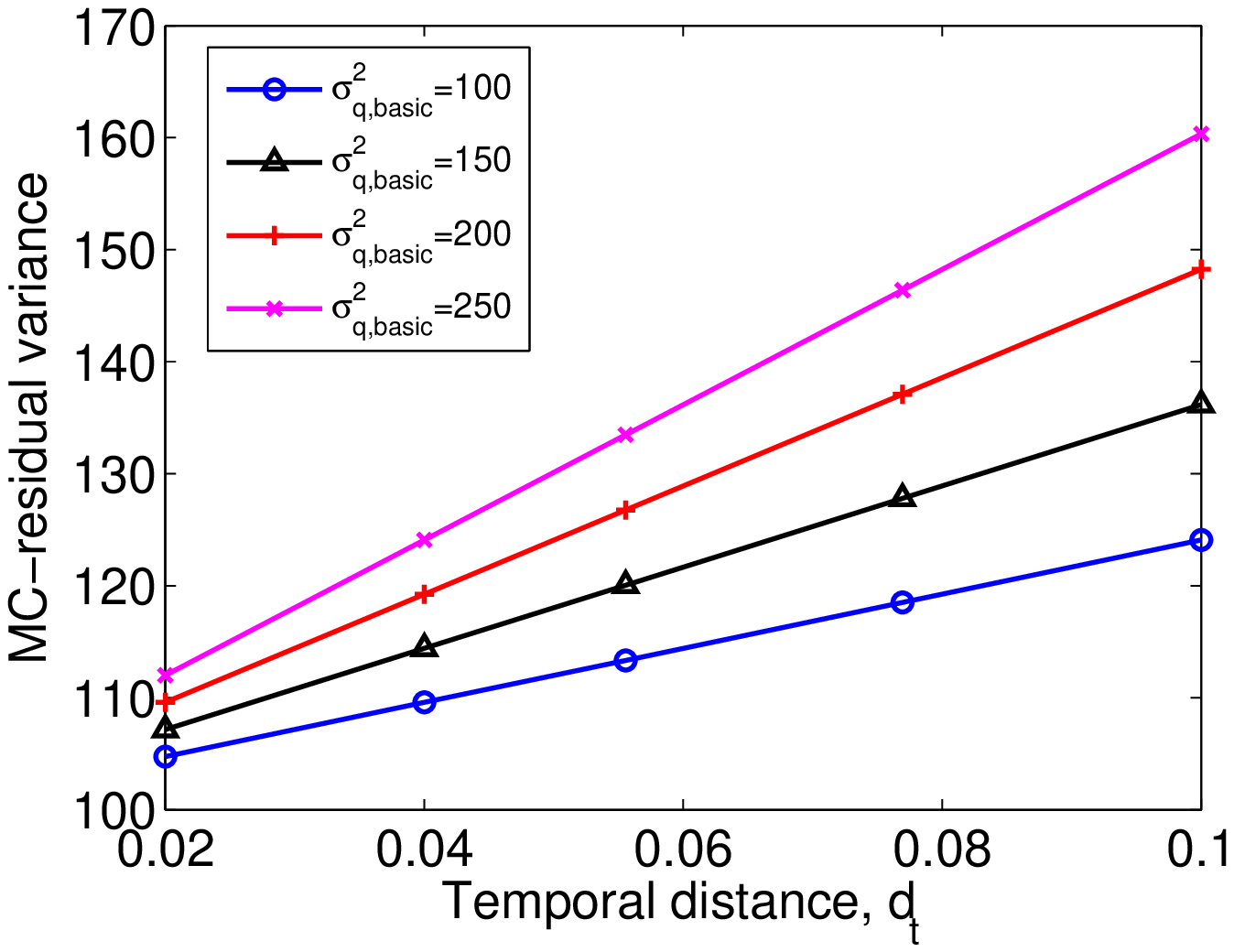}
\label{Fig:MC_coding_residual_variance_estimation_vs_temporal_distance_for_various_q_var_basic}}
\caption{Estimation of MC-residual variance in MC-coding. (a) as function of bit-rate for various frame-rates (temporal distances). (b) as function of temporal-distance for various bit-rates. (c) as function of temporal-distance for various motion-energy values $\sigma ^2 _{q,basic}$.}
\label{Fig:Estimation of MC-residual variance in MC-coding}
\end{figure*}

First, we examine the estimated variance as the bit-rate varies (Fig. \ref{Fig:MC_coding_residual_variance_estimation_vs_bpp_for_various_framerates}). The variance is monotonically decreasing as the bit-rate increases, this is due to improved quality of the reference frame that increases its similarity to the coded frame. The graphs have a convex shape as expected from a distortion-rate function.

The estimated variance as the frame-rate varies is presented in (Fig. \ref{Fig:MC_coding_residual_variance_estimation_vs_temporal_distance_for_various_bpps}).
We assume the reference and the coded frames are adjacent, hence the frame-rate and the temporal-distance can be alternately referred using $d_t = \frac{1}{F_{rate}}$.
The variance is linearly increasing as the temporal-distance increases. This is justified by the reduced similarity between the reference and coded frames as they get farther.

We compared our estimation for varying motion-complexity of the coded video expressed by $\sigma ^2 _{q,basic}$ (Fig. \ref{Fig:MC_coding_residual_variance_estimation_vs_temporal_distance_for_various_q_var_basic}).
The estimated variance increases together with the motion-complexity. This conforms with the fact that more complex motion affects the motion-estimation results and increases the MC-residual energy.

\subsection{Motion-Compensated Frame-Rate Up Conversion}
\begin{figure*}[!t]
\centering
{\subfloat[]{\includegraphics[width=2.3in]{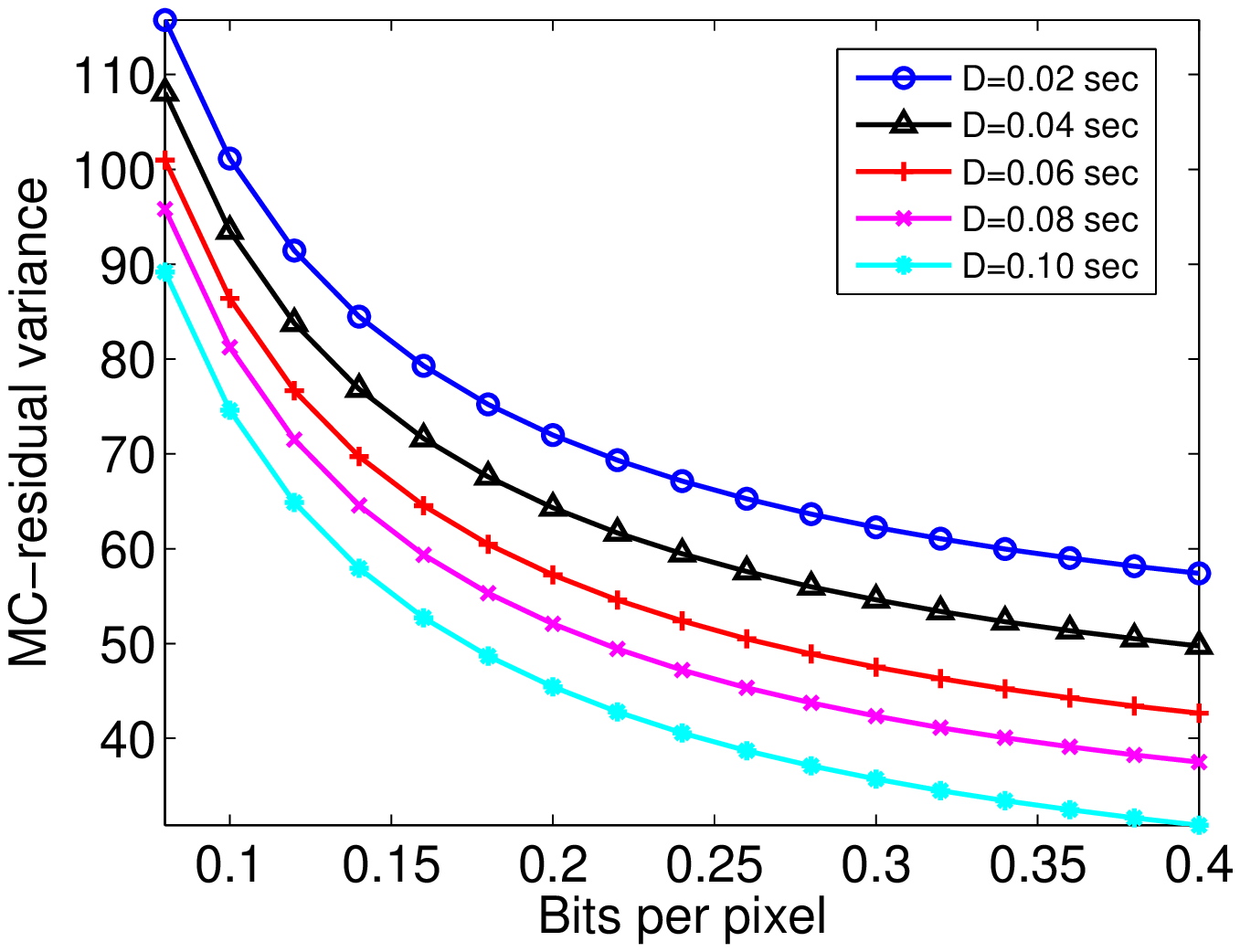}
\label{Fig:MC_FRUC_residual_variance_estimation_vs_bpp_for_various_framerates}}}
\subfloat[]{\includegraphics[width=2.3in]{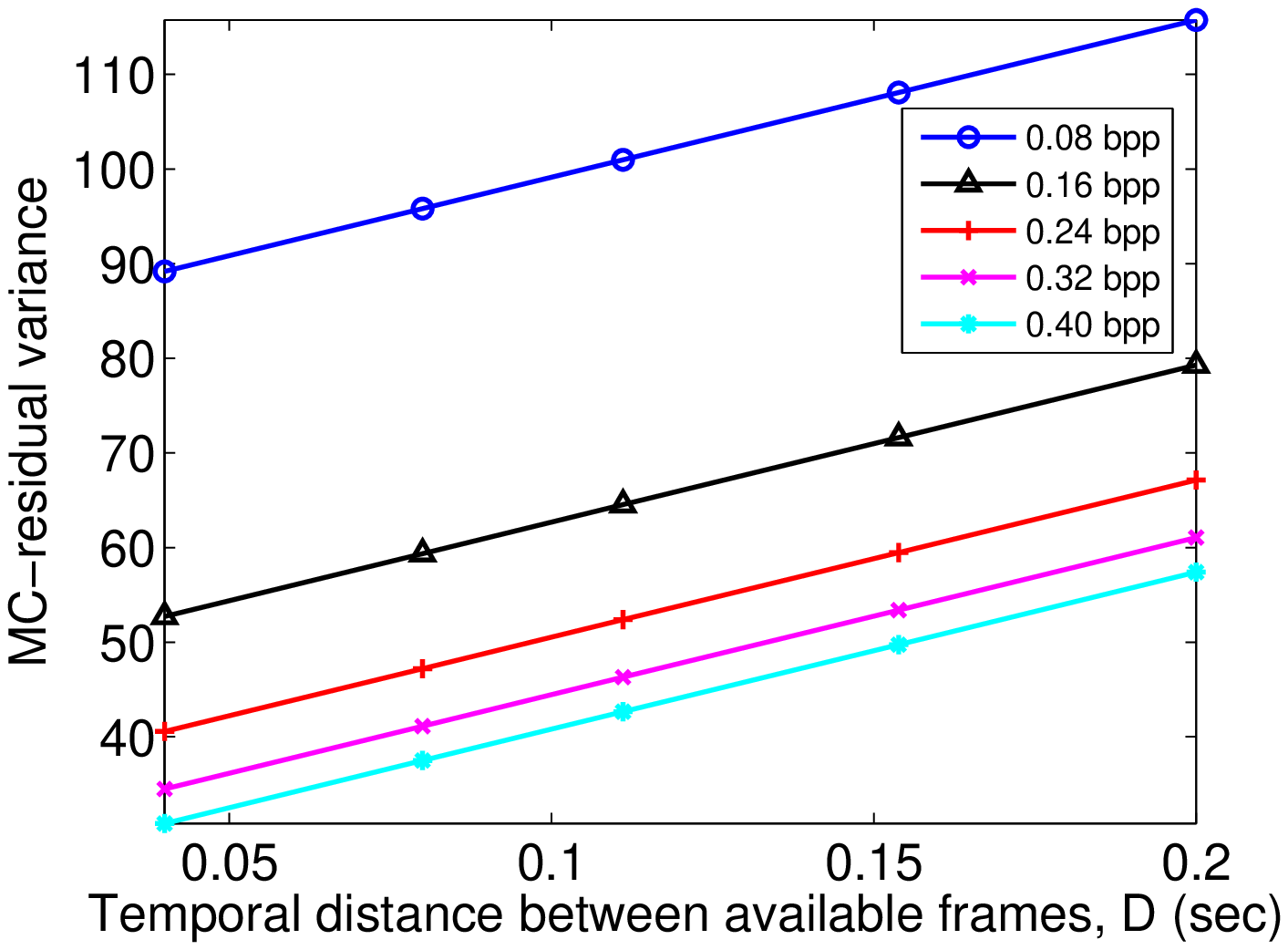}
\label{Fig:MC_FRUC_residual_variance_estimation_vs_temporal_distance_for_various_bpps}}
\subfloat[]{\includegraphics[width=2.3in]{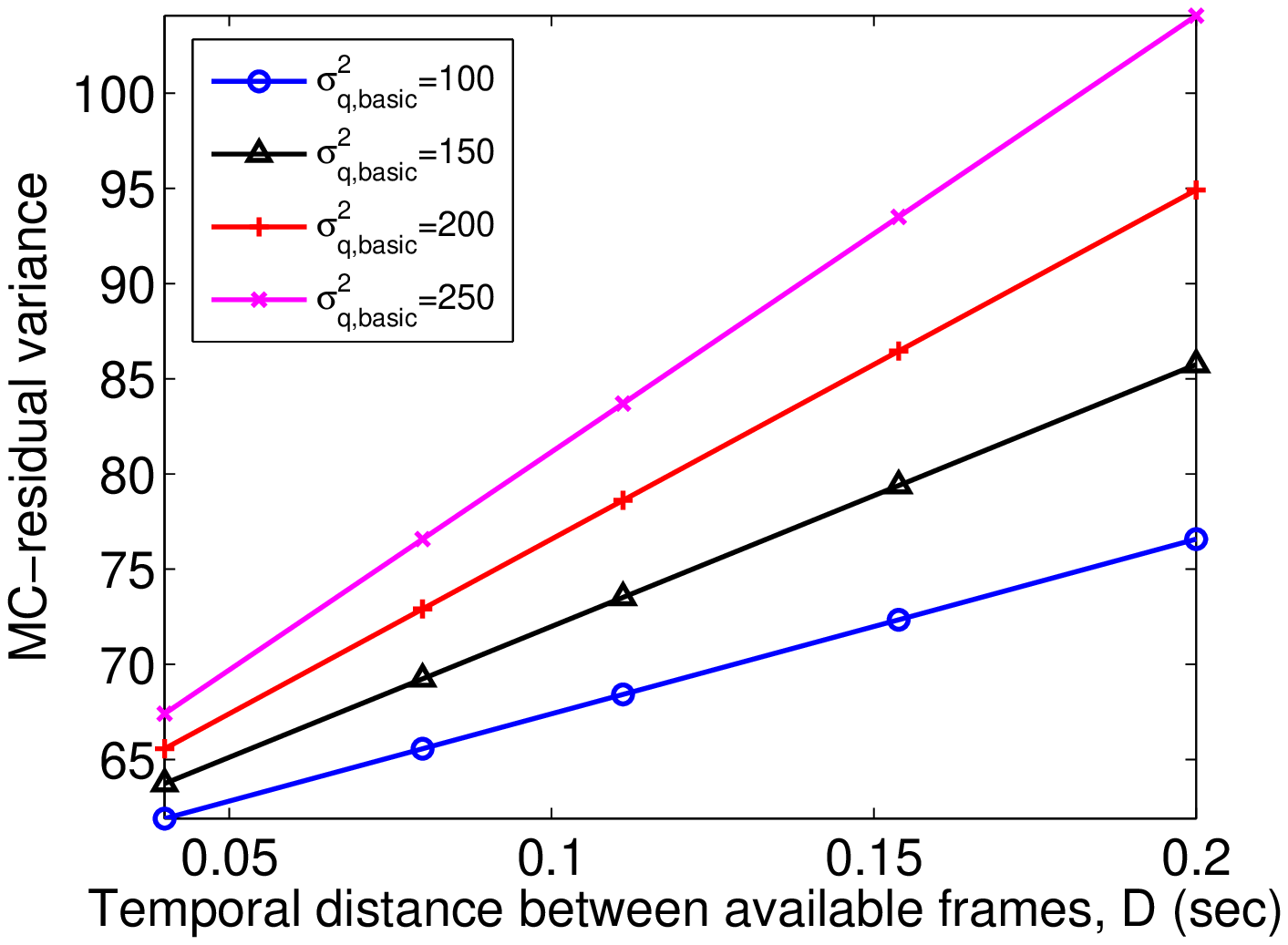}
\label{Fig:MC_FRUC_residual_variance_estimation_vs_temporal_distance_for_various_q_var_basic}}
\caption{Estimation of MC-residual variance in MC-FRUC (i.e., estimation of interpolation MSE). $\gamma _{abs}=2$. (a) as function of bit-rate for various interpolation factors (temporal distances). (b) as function of interpolation factors (temporal-distance) for various bit-rates. (c) as function of temporal-distance for various motion-energy values $\sigma ^2 _{q,basic}$.}
\label{Fig:Estimation of MC-residual variance in MC-FRUC}
\end{figure*}

Let us consider our estimations for the MC-FRUC MSE (\ref{eq:MC-prediction - absent frame - error variance - non static region}), (\ref{eq:MC-prediction - absent frame - error variance - non static region - for theta 0.5}). The equations for the MC-FRUC MSE (\ref{eq:MC-prediction - absent frame - error variance - non static region - for theta 0.5}) and the residual variance in MC-coding (\ref{eq:MC-prediction - available frame error variance - temporal distance in seconds}) are similar; therefore, similar behavior is expected. The estimations (Fig. \ref{Fig:Estimation of MC-residual variance in MC-FRUC}) conform with these expectations. The explanations given above for MC-coding (see section \ref{subsec:Theoretical Estimations - Motion-Compensated Coding}) also hold here.


\section{Experimental Results}
\label{sec:Experimental Results}
\subsection{Motion-Compensated Coding}
\begin{figure*}[!t]
\centering
{\subfloat[]{\includegraphics[width=3in]{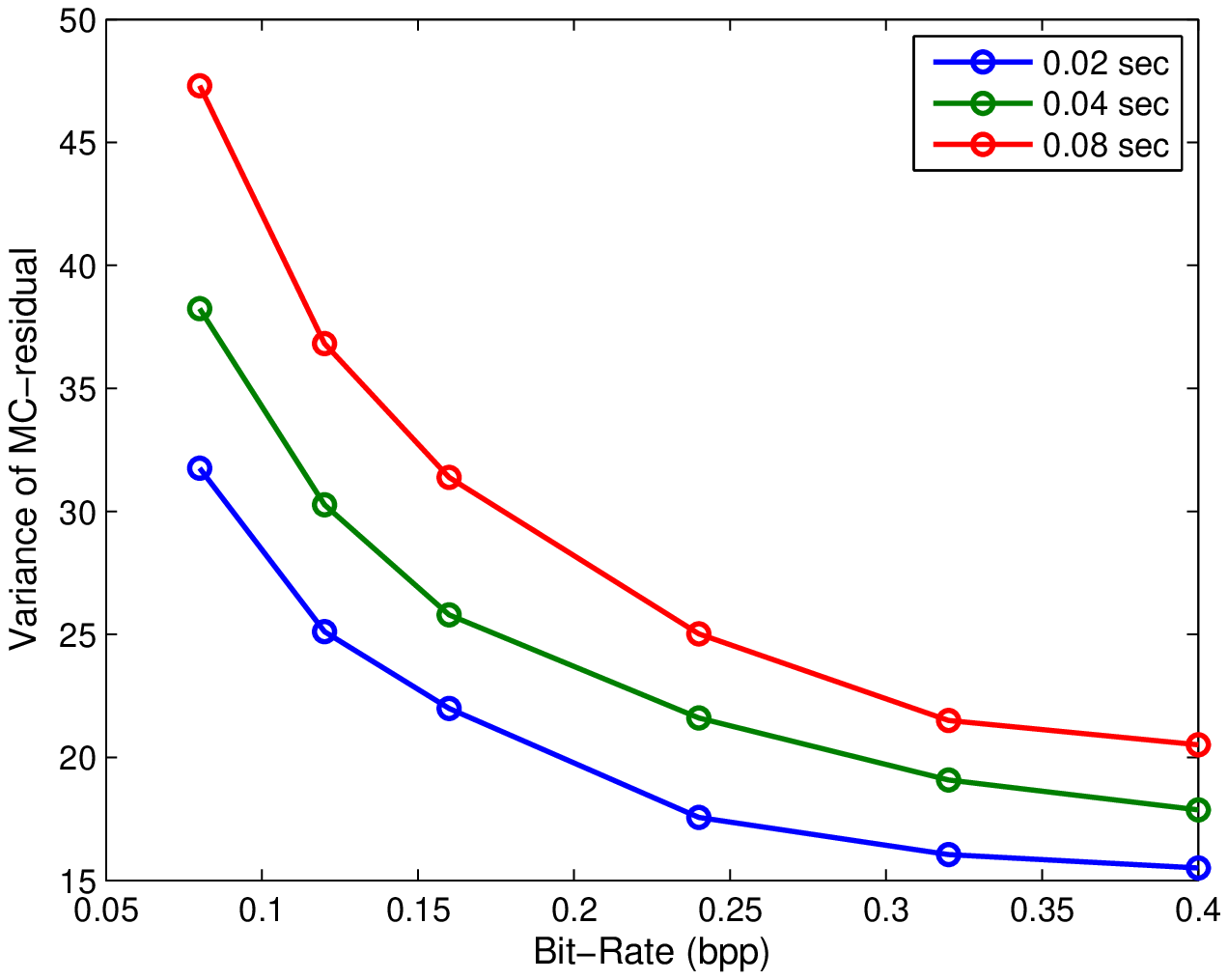}%
\label{Fig:MC_coding_residual_vs_bitrate_experimental_results_oldtowncross10sec}}}
\subfloat[]{\includegraphics[width=3in]{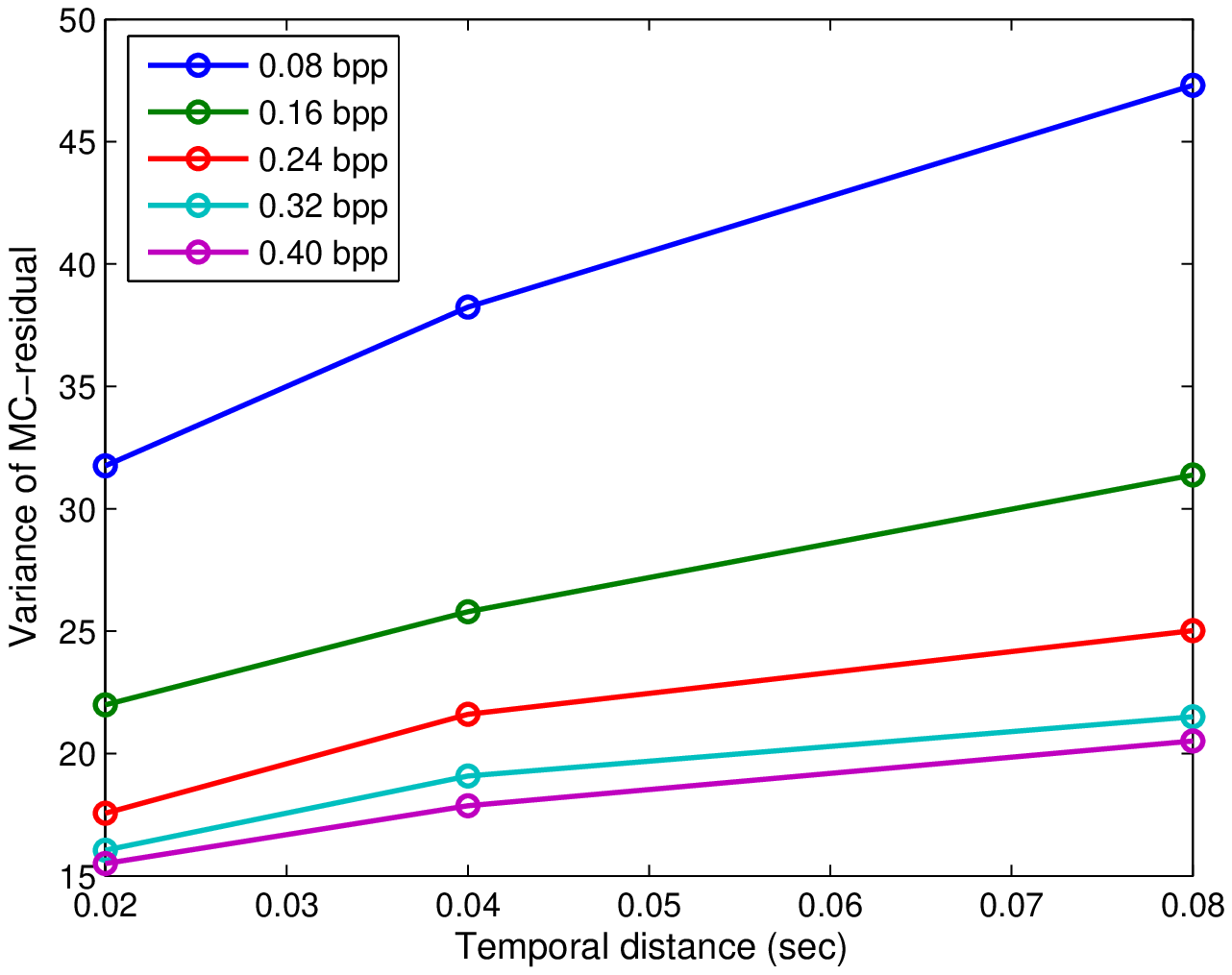}%
\label{Fig:MC_coding_residual_vs_temporal_distance_experimental_results_oldtowncross10sec}}
\caption{Measured MC-residual statistics in MC-coding of 'Old town cross' sequence (grayscale, frame size 720x720, 10 seconds length). (a) as function of bit-rate for various temporal-distance values. (b) as function of temporal-distance for various bit-rates.}
\label{Fig:Old town cross experimental results of MC-residual statistics in MC-coding}
\end{figure*}

\begin{figure*}[!t]
\centering
{\subfloat[]{\includegraphics[width=3in]{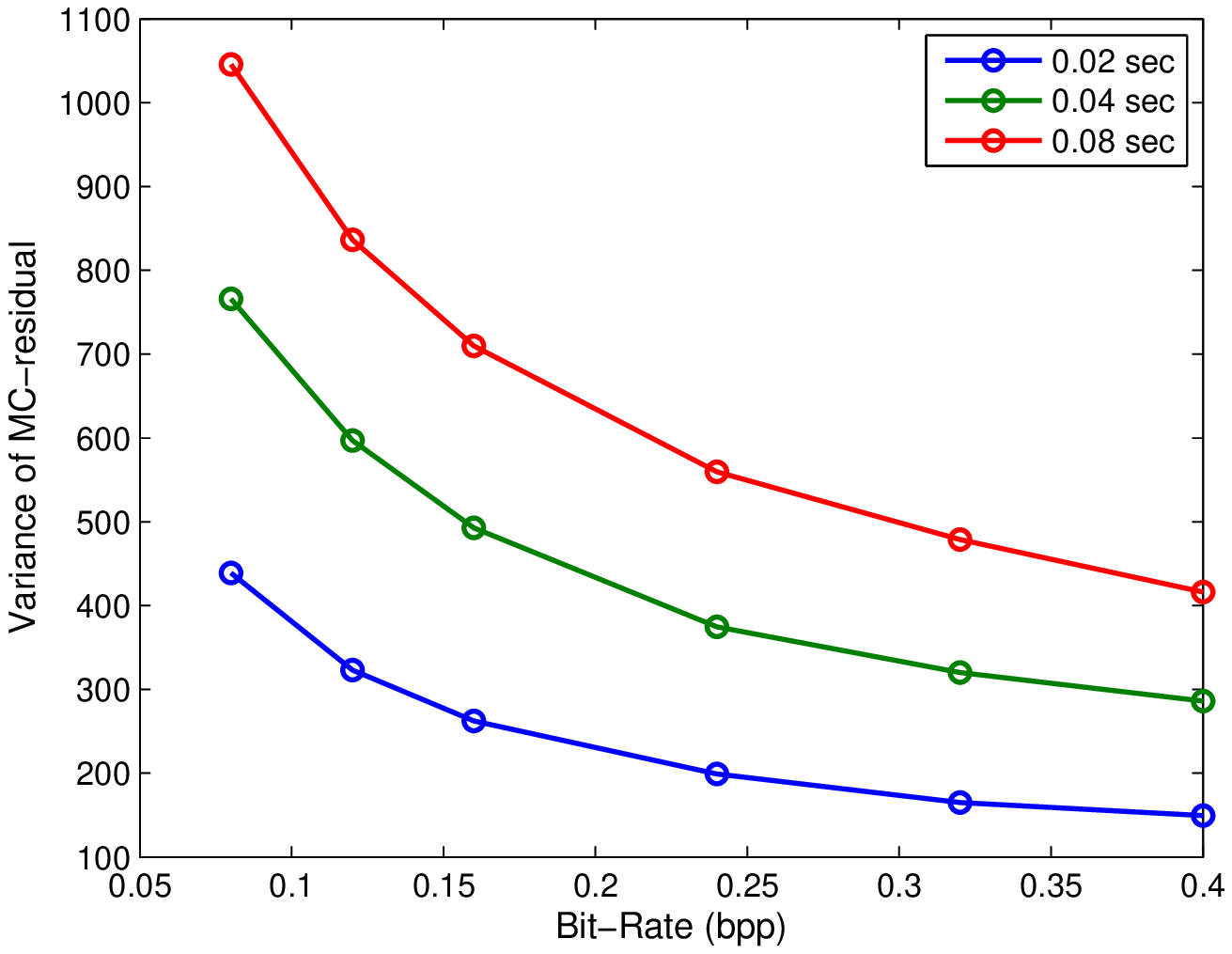}%
\label{Fig:MC_coding_residual_vs_bitrate_experimental_results_parkrun8sec}}}
\subfloat[]{\includegraphics[width=3in]{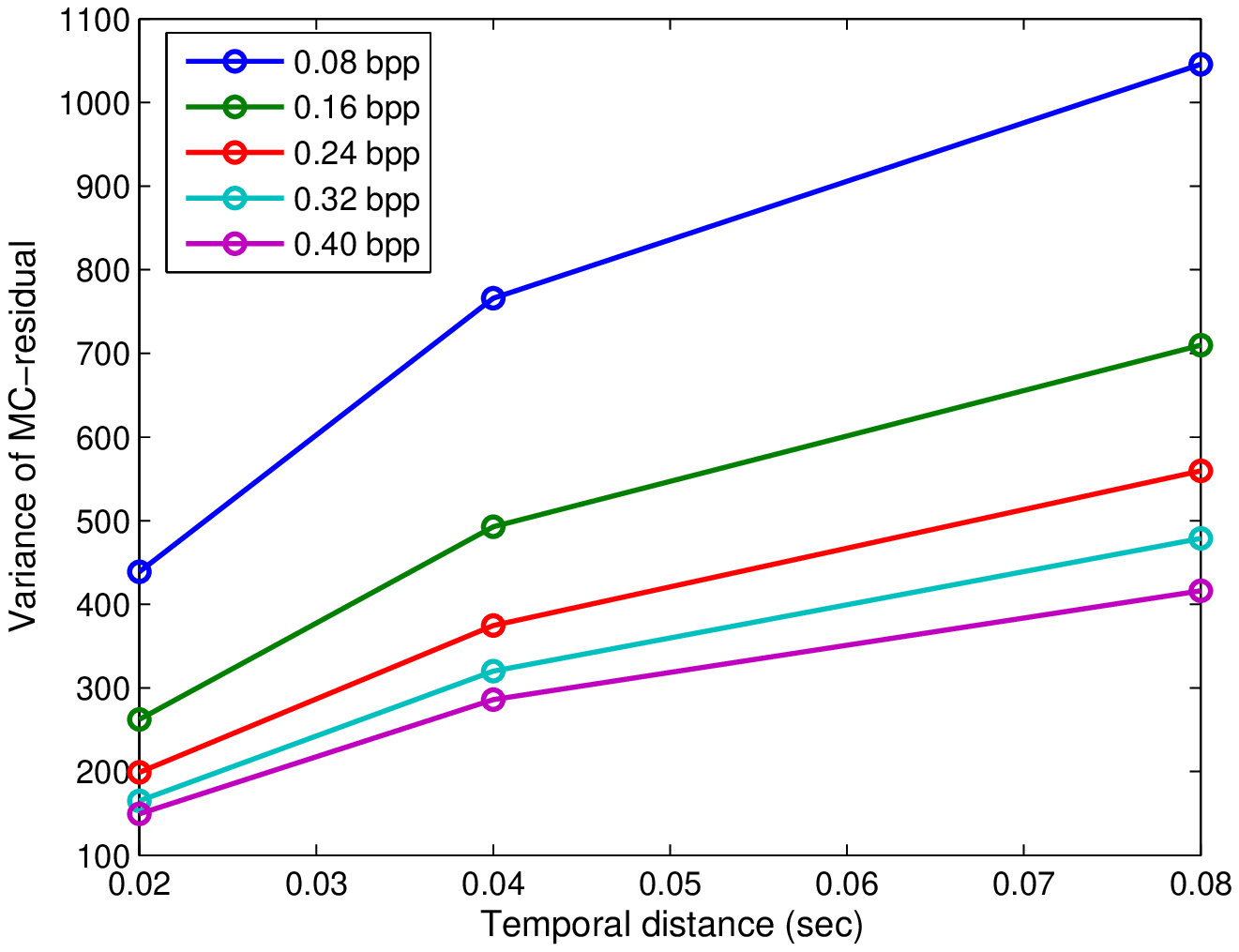}%
\label{Fig:MC_coding_residual_vs_temporal_distance_experimental_results_parkrun8sec}}
\caption{Measured MC-residual statistics in MC-coding of 'Parkrun' sequence (grayscale, frame size 720x720, 8 seconds length). (a) as function of bit-rate for various temporal-distance values. (b) as function of temporal-distance for various bit-rates.}
\label{Fig:Parkrun experimental results of MC-residual statistics in MC-coding}
\end{figure*}

We measured the average MC-residual variance in an H.264 software \cite{RefWorks:72} for the 'old town cross' and 'Parkrun' sequences (Fig. \ref{Fig:Old town cross experimental results of MC-residual statistics in MC-coding},\ref{Fig:Parkrun experimental results of MC-residual statistics in MC-coding}). The variance has a monotonically decreasing convex shape as function of the bit-rate (Figs. \ref{Fig:MC_coding_residual_vs_bitrate_experimental_results_oldtowncross10sec}, \ref{Fig:MC_coding_residual_vs_bitrate_experimental_results_parkrun8sec}), as in our model (Fig. \ref{Fig:MC_coding_residual_variance_estimation_vs_bpp_for_various_framerates}). In addition, the variance has a relatively linearly-increasing behavior as function of the temporal-distance (Figs. \ref{Fig:MC_coding_residual_vs_temporal_distance_experimental_results_oldtowncross10sec}, \ref{Fig:MC_coding_residual_vs_temporal_distance_experimental_results_parkrun8sec}), this also conforms with our model estimations (Fig. \ref{Fig:MC_coding_residual_variance_estimation_vs_temporal_distance_for_various_bpps}). The 'parkrun' sequence contains more complex motion than 'old town cross'; as a result, its residual variance values are significantly higher (Figs. \ref{Fig:MC_coding_residual_vs_bitrate_experimental_results_oldtowncross10sec}, \ref{Fig:MC_coding_residual_vs_bitrate_experimental_results_parkrun8sec}). This is also expressed in our model (Fig. \ref{Fig:MC_coding_residual_variance_estimation_vs_temporal_distance_for_various_q_var_basic}).

\subsection{Motion-Compensated Frame-Rate Up Conversion}
\begin{figure*}
\centering
{\subfloat[]{\includegraphics[width=3in]{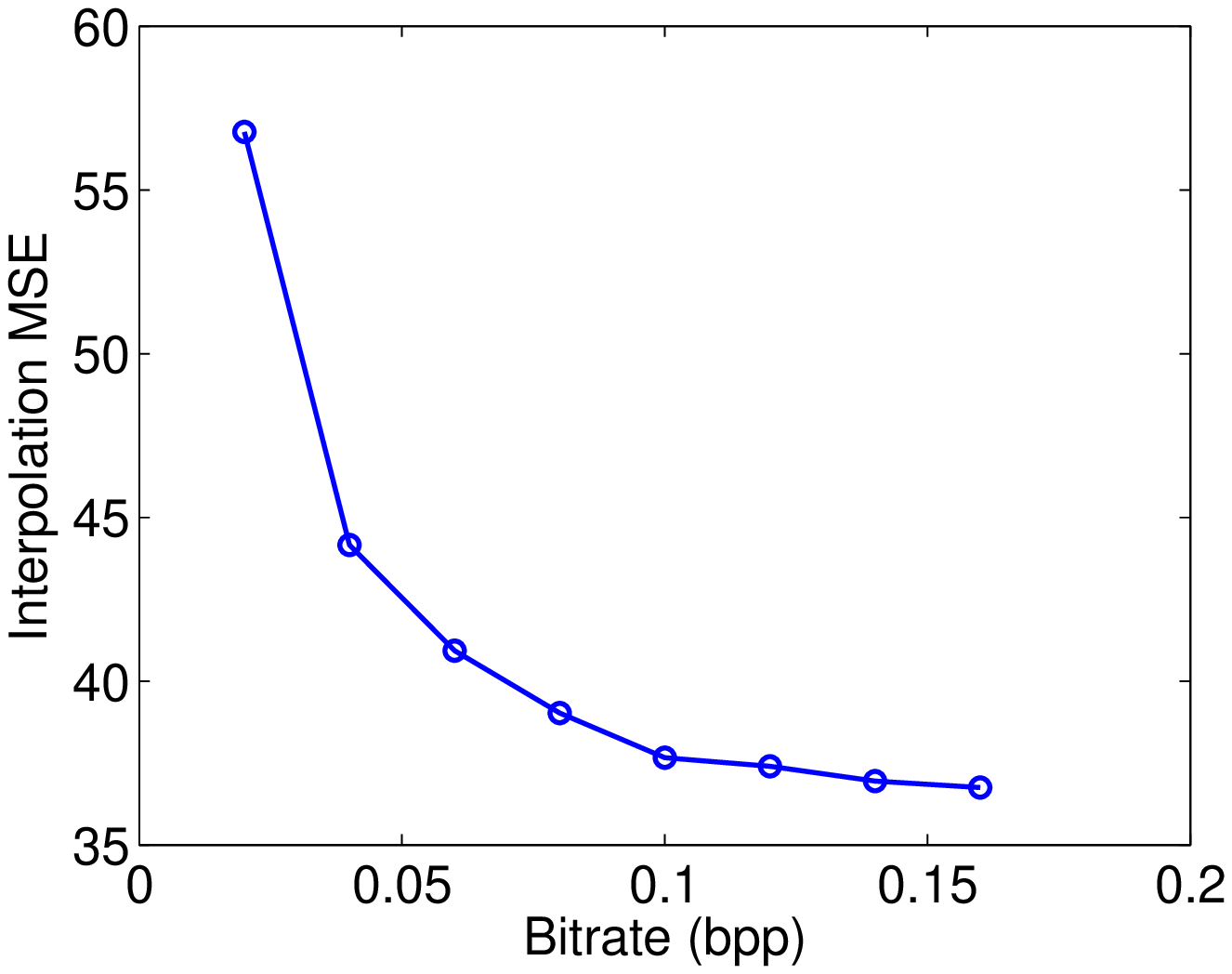}%
\label{Fig:MC_FRUC_MSE_experimental_results_ice4cif_vs_bitrate}}}
\subfloat[]{\includegraphics[width=3in]{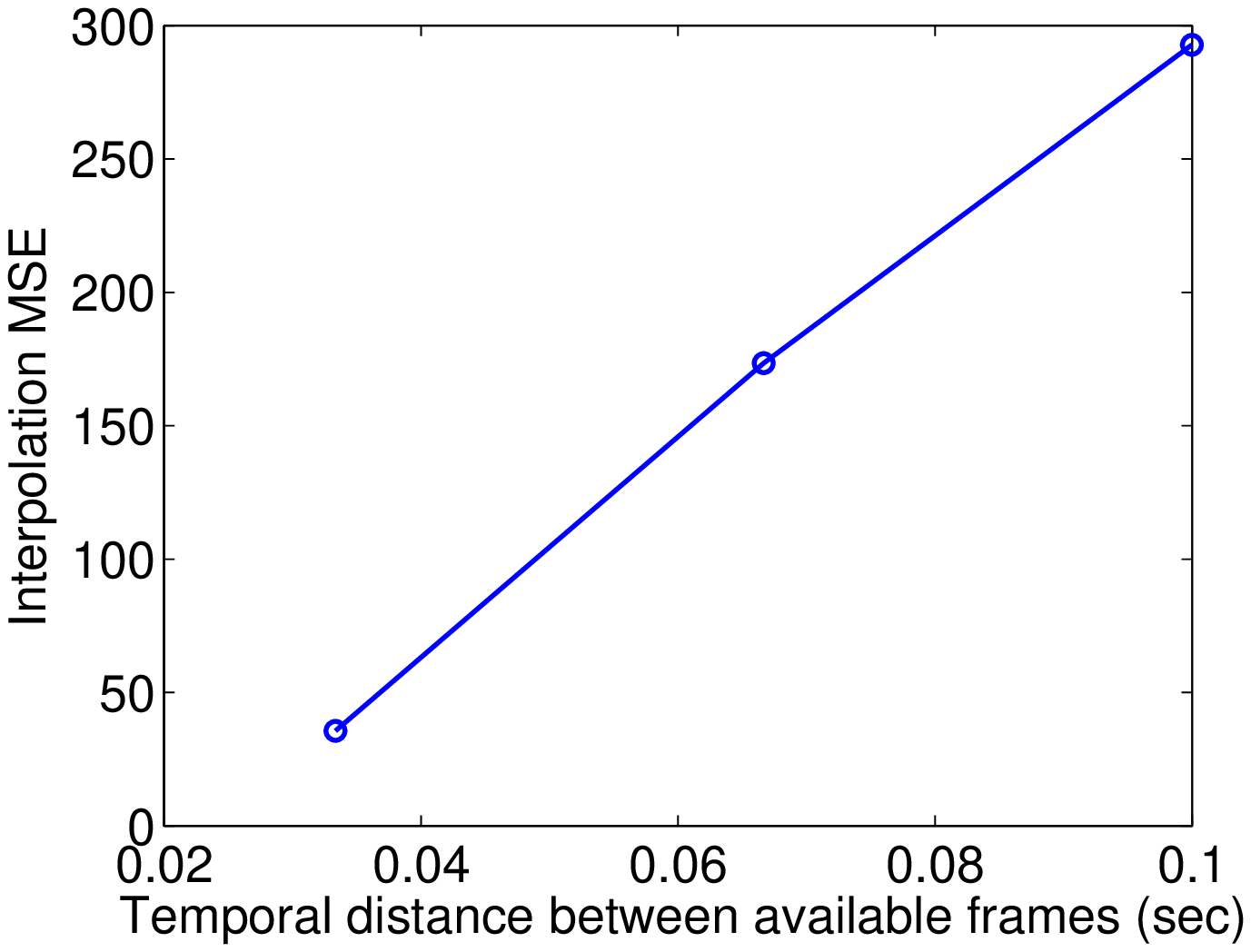}%
\label{Fig:MC_FRUC_MSE_experimental_results_ice4cif_vs_temporal_distance_no_compression}}
\caption{Measured MSE in MC-FRUC applied on 'Ice' sequence (grayscale, frame size 576x576, 60fps). (a) as function of bit-rate. (b) as function of temporal-distance (i.e., varying temporal-interpolation factors) for raw video.}
\label{Fig:Ice experimental results of MSE in MC-FRUC}
\end{figure*}

\begin{figure*}
\centering
{\subfloat[]{\includegraphics[width=3in]{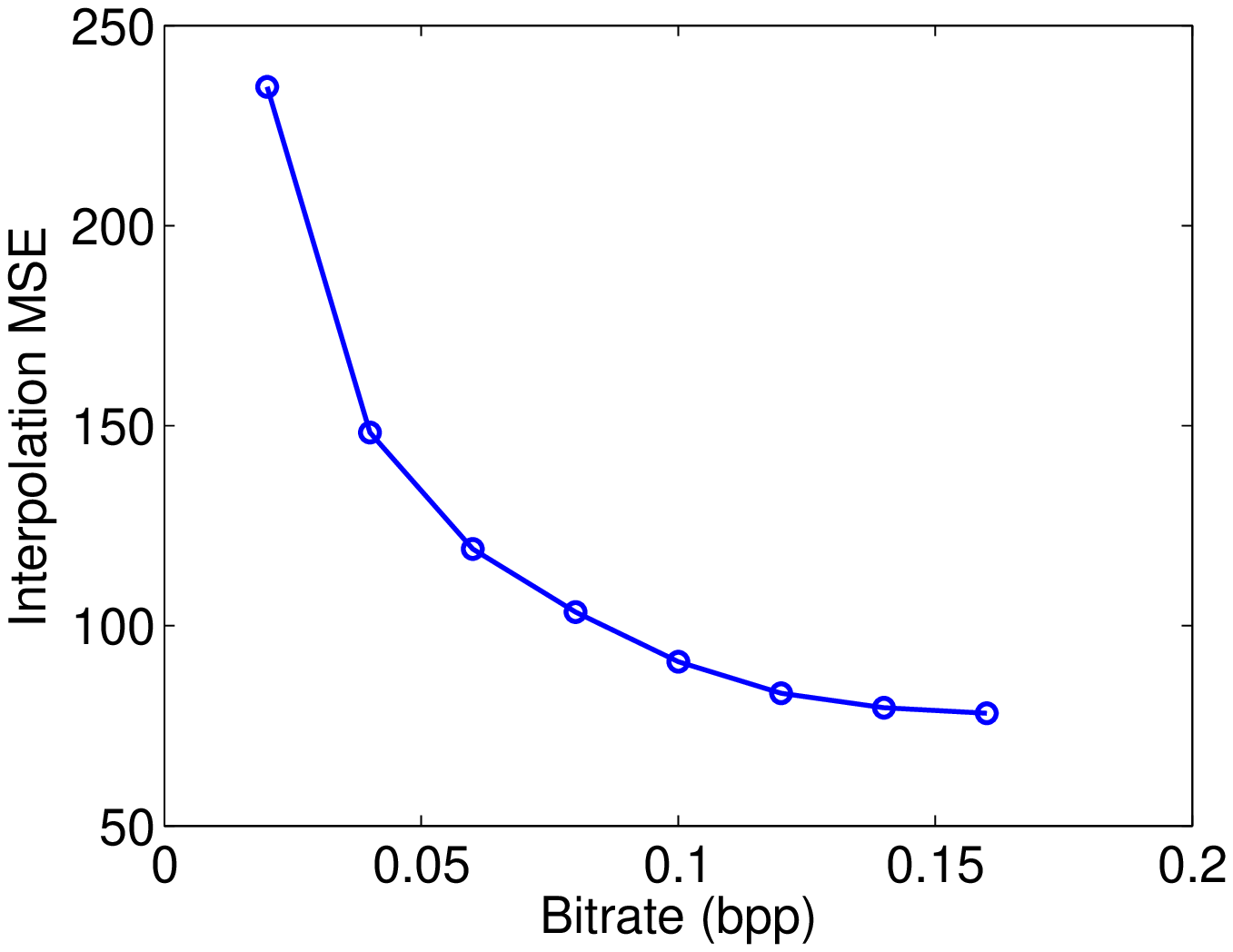}%
\label{Fig:MC_FRUC_MSE_experimental_results_harbour4cif_vs_bitrate}}}
\subfloat[]{\includegraphics[width=3in]{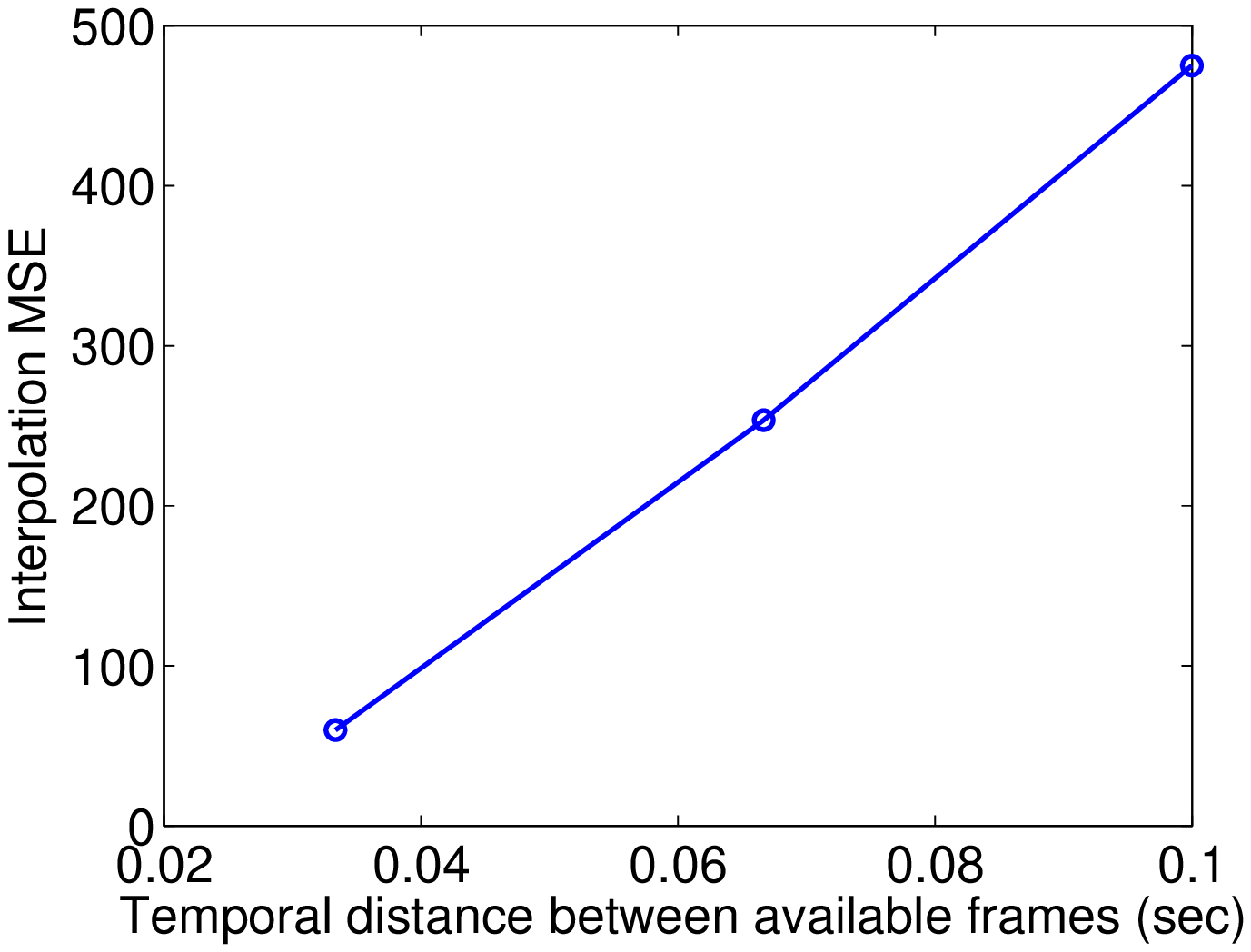}%
\label{Fig:MC_FRUC_MSE_experimental_results_harbour4cif_vs_temporal_distance_no_compression}}
\caption{Measured MSE in MC-FRUC applied on 'Harbour' sequence (grayscale, frame size 576x576, 60fps). (a) as function of bit-rate. (b) as function of temporal-distance (i.e., varying temporal-interpolation factors) for raw video.}
\label{Fig:Harbour experimental results of MSE in MC-FRUC}
\end{figure*}

\begin{figure*}
\centering
{\subfloat[]{\includegraphics[width=3in]{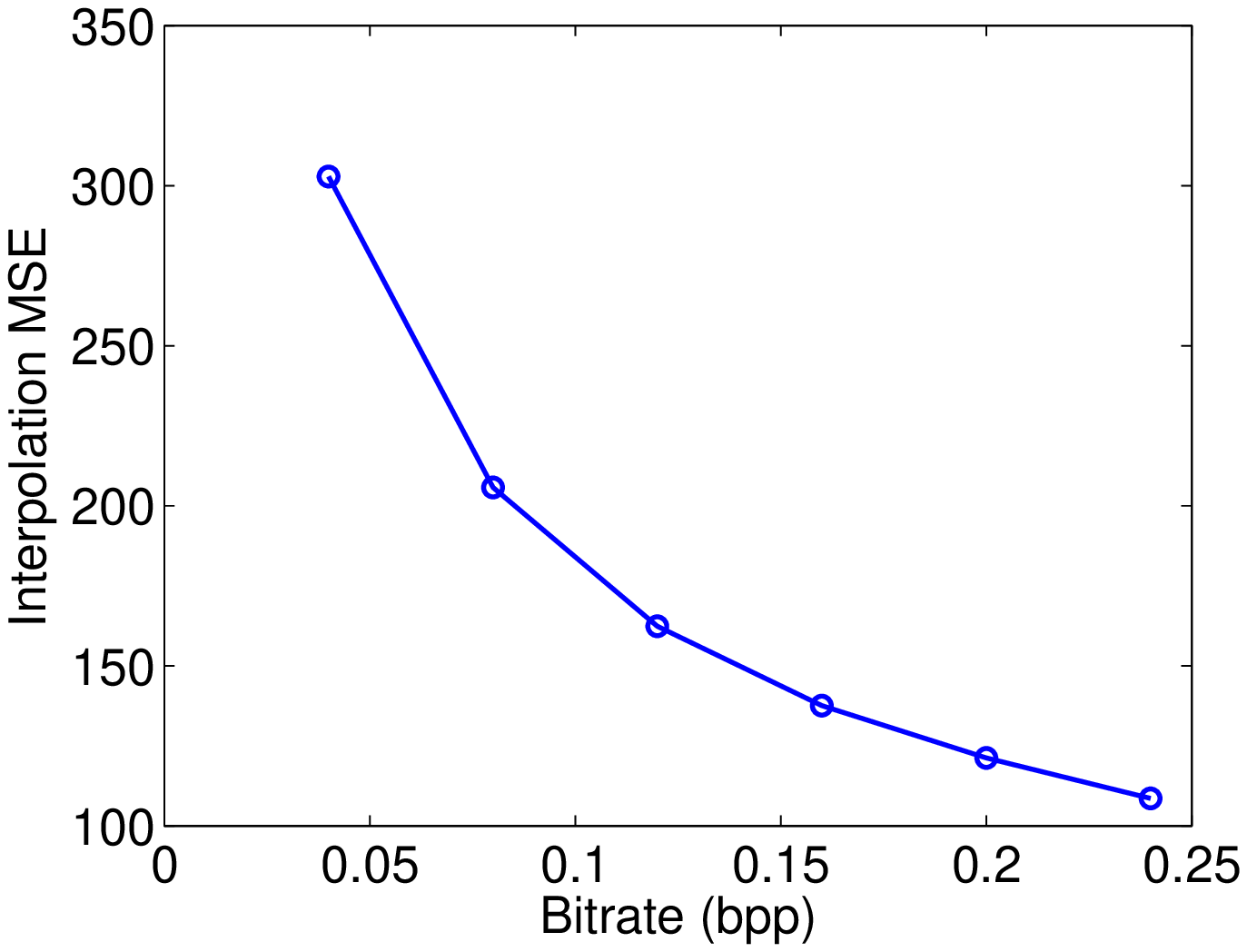}%
\label{Fig:MC_FRUC_MSE_experimental_results_parkrun720x720_vs_bitrate}}}
\subfloat[]{\includegraphics[width=3in]{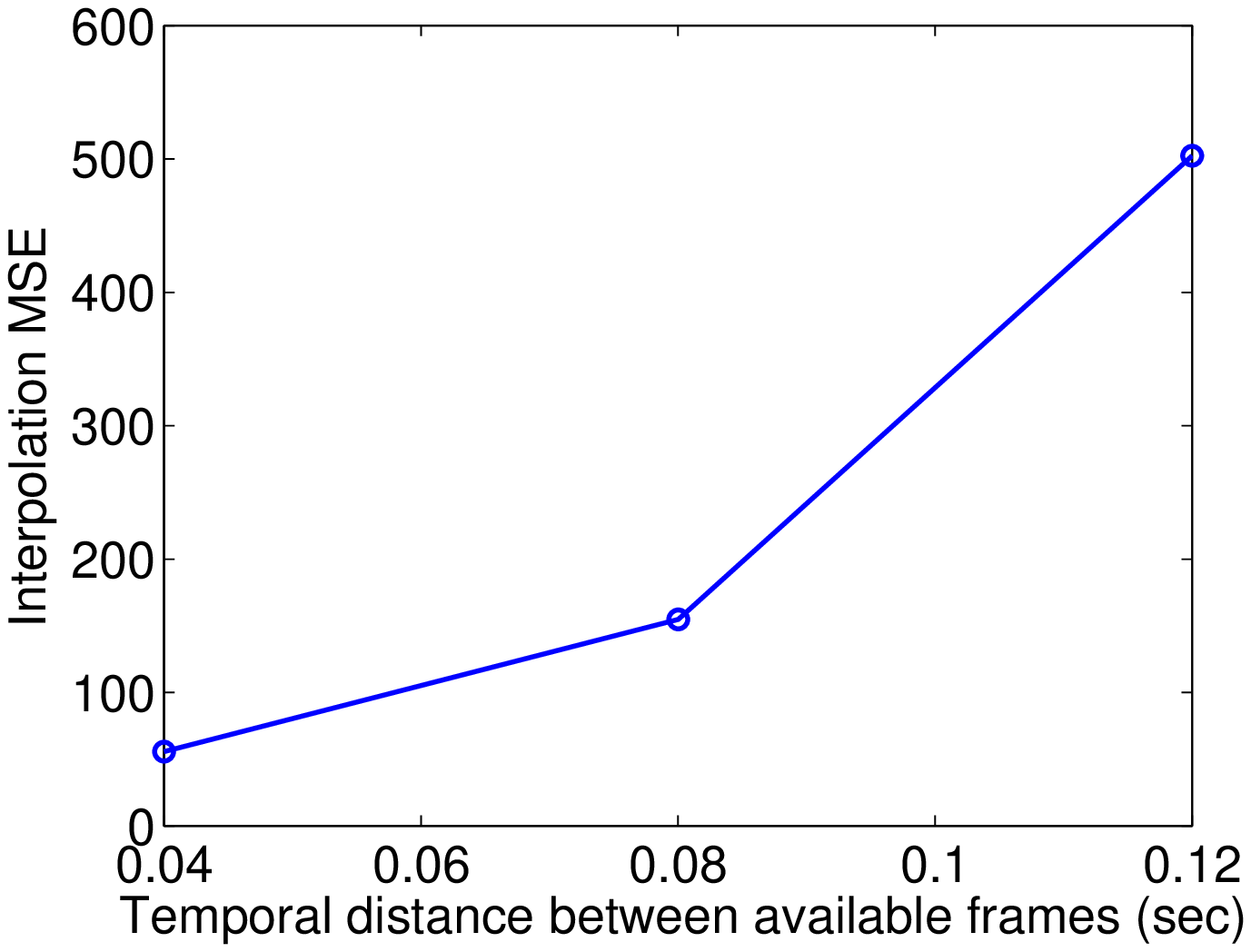}%
\label{Fig:MC_FRUC_MSE_experimental_results_parkrun720x720_vs_temporal_distance_no_compression}}
\caption{Measured MSE in MC-FRUC applied on 'Parkrun' sequence (grayscale, frame size 720x720, 50fps). (a) as function of bit-rate. (b) as function of temporal-distance (i.e., varying temporal-interpolation factors) for raw video.}
\label{Fig:Parkrun experimental results of MSE in MC-FRUC}
\end{figure*}

In section \ref{subsec:MC-Prediction of an Absent Frame} we gave an expression for the MC-FRUC error (\ref{eq:MC-prediction - absent frame - error variance - non static region - for theta 0.5}). Here we compare the behavior of the theoretical model with experimental results obtained from an MC-FRUC procedure implemented in Matlab. The variance of MC-prediction error in FRUC equals to the interpolation MSE; hence, we refer them here interchangeably. 
We examined the dependency of FRUC MSE in temporal-distance and bit-rate.
For our experiments, we implemented an MC-FRUC algorithm that applies bidirectional motion-estimation with half-pel accuracy. We considered the central-interpolated frames for upsampling factors $D=2,4,6$ (i.e., $j=\frac{D}{2}$ for even $D$ values). Hence, we studied the relation of the MSE to the temporal-distance by applying FRUC at a varying interpolation factor, $D$, for a fixed frame-rate.
The experiments showed an approximately linear increment of the MSE together with the temporal-distance (Figs. \ref{Fig:MC_FRUC_MSE_experimental_results_ice4cif_vs_temporal_distance_no_compression},\ref{Fig:MC_FRUC_MSE_experimental_results_harbour4cif_vs_temporal_distance_no_compression},\ref{Fig:MC_FRUC_MSE_experimental_results_parkrun720x720_vs_temporal_distance_no_compression}). In addition, its relation to the bit-rate has a convex-decreasing shape (Figs. \ref{Fig:MC_FRUC_MSE_experimental_results_ice4cif_vs_bitrate},\ref{Fig:MC_FRUC_MSE_experimental_results_harbour4cif_vs_bitrate},\ref{Fig:MC_FRUC_MSE_experimental_results_parkrun720x720_vs_bitrate}). 
'Ice' sequence contains more static regions than 'Harbour', i.e., its motion is simpler. Accordingly, higher MSE values are observed for 'Harbour'.
The above observations are expressed correspondingly in the theoretical estimations (Fig. \ref{Fig:Estimation of MC-residual variance in MC-FRUC}).

\section{Conclusion}
\label{sec:Conclusion}
The motion-compensation procedure was studied in this work. 
Both cases of predicting available and absent frames were theoretically examined, and expressions for the prediction error and its autocorrelation were given. The considered procedures represent the applications of MC in coding and FRUC.
The analysis is based on a statistical model for the video signal that was presented in the beginning of this paper. Along this study, a special focus was given to the effects of frame-rate and bit-rate on the MC-prediction error.
The MC applications in coding and FRUC were studied in the same theoretic framework. Hence, this paper can be seen as a comparison between the applications, as the similarities and differences raise from the text.
For the application of MC-coding, we presented three autocorrelation models at different levels of analytic simplicity. Analytic simplicity is useful for examination of complex systems that include MC-coding. Future work can analyze such systems.
This work emphasizes the significant effect of frame-rate and bit-rate on MC performance. Future work can suggest MC-related algorithms that consider these factors adaptively.


%

\clearpage
\appendix

\section{Autocorrelation Calculation for Prediction of an Available Frame}
\label{appendix_sec:MC Prediction of an Available Frame}
Here we calculate the autocorrelation of the MC-prediction residual for the case of an available frame that was presented in section \ref{sec:MC-Prediction of an Available Frame}. 
Recall the expression given in (\ref{eq:MC-prediction - frame prediction model}) for the MC-prediction:
\begin{IEEEeqnarray}{rCl}
\label{eq:MC-prediction - frame prediction model - appendix}
\nonumber && {{\hat f}_t}\left( {x,y\left| {{f_{t - i}^{ref}},} \right.\hat \varphi \left( {t,t - i\left| {{f_t},{f_{t - i}}} \right.} \right)} \right) = v\left( {x - {\varphi _x}\left( {t,0} \right) - \Delta x,y - {\varphi _y}\left( {t,0} \right) - \Delta y} \right)
\\ 
&& \qquad {} + {n_{t - i}^{ref}}\left( {x - {{\hat \varphi }_x}\left( {t,t - i\left| {{f_t},{f_{t - i}}} \right.} \right),y - {{\hat \varphi }_y}\left( {t,t - i\left| {{f_t},{f_{t - i}}} \right.} \right)} \right)
\end{IEEEeqnarray}

Let us assume that $\Delta x$ and $\Delta y$ are small. Then, first-order Taylor expansion gives the following approximations
\begin{IEEEeqnarray}{rCl}
\label{eq:MC-prediction - v image Taylor expansion}
v\left( {x - {\varphi _x}\left( {t,0} \right) - \Delta x,y - {\varphi _y}\left( {t,0} \right) - \Delta y} \right) & \approx & v\left( {x - {\varphi _x}\left( {t,0} \right),y - {\varphi _y}\left( {t,0} \right)} \right)
\\ \nonumber
&& - \Delta x{\left. {\frac{\partial }{{\partial \tilde x}}v\left( {\tilde x,\tilde y} \right)} \right|_{\left( {\tilde x,\tilde y} \right) = \left( {x - {\varphi _x}\left( {t,0} \right),y - {\varphi _y}\left( {t,0} \right)} \right)}}
\\ \nonumber
&& - \Delta y{\left. {\frac{\partial }{{\partial \tilde y}}v\left( {\tilde x,\tilde y} \right)} \right|_{\left( {\tilde x,\tilde y} \right) = \left( {x - {\varphi _x}\left( {t,0} \right),y - {\varphi _y}\left( {t,0} \right)} \right)}}
\end{IEEEeqnarray}
\begin{IEEEeqnarray}{rCl}
\label{eq:MC-prediction - noise image Taylor expansion}
&& {n_{t - i}}\left( {x - {{\hat \varphi }_x}\left( {t,t - i\left| {{f_t},{f_{t - i}}} \right.} \right),y - {{\hat \varphi }_y}\left( {t,t - i\left| {{f_t},{f_{t - i}}} \right.} \right)} \right) \approx 
\\ \nonumber
&& \qquad {} {n_{t - i}}\left( {x - {\varphi _x}\left( {t,t - i} \right),y - {\varphi _y}\left( {t,t - i} \right)} \right)
\\ \nonumber
&& \qquad {}  - \Delta x{\left. {\frac{\partial }{{\partial \tilde x}}{n_{t - i}}\left( {\tilde x,\tilde y} \right)} \right|_{\left( {\tilde x,\tilde y} \right) = \left( {x - {\varphi _x}\left( {t,t - i} \right),y - {\varphi _y}\left( {t,t - i} \right)} \right)}}
\\ \nonumber
&& \qquad {}   - \Delta y{\left. {\frac{\partial }{{\partial \tilde y}}{n_{t - i}}\left( {\tilde x,\tilde y} \right)} \right|_{\left( {\tilde x,\tilde y} \right) = \left( {x - {\varphi _x}\left( {t,t - i} \right),y - {\varphi _y}\left( {t,t - i} \right)} \right)}}
\end{IEEEeqnarray}
Substitution of (\ref{eq:MC-prediction - v image Taylor expansion}) and (\ref{eq:MC-prediction - noise image Taylor expansion}) into (\ref{eq:MC-prediction - frame prediction model}) yields
\begin{IEEEeqnarray}{rCl}
\label{eq:MC-prediction - approximated motion estimation model}
{{\hat f}_t}\left( {x,y\left| {{f_{t - i}^{ref}},} \right.\hat \varphi \left( {t,t - i\left| {{f_t},{f_{t - i}}} \right.} \right)} \right) & = & v\left( {x - {\varphi _x}\left( {t,0} \right),y - {\varphi _y}\left( {t,0} \right)} \right)
\\ \nonumber
&& + {n_{t - i}^{ref}}\left( {x - {\varphi _x}\left( {t,t - i} \right),y - {\varphi _y}\left( {t,t - i} \right)} \right)
\\ \nonumber
&& - \Delta x{\left. {\frac{\partial }{{\partial \tilde x}}v\left( {\tilde x,\tilde y} \right)} \right|_{\left( {\tilde x,\tilde y} \right) = \left( {x - {\varphi _x}\left( {t,0} \right),y - {\varphi _y}\left( {t,0} \right)} \right)}}
\\ \nonumber
&& - \Delta y{\left. {\frac{\partial }{{\partial \tilde y}}v\left( {\tilde x,\tilde y} \right)} \right|_{\left( {\tilde x,\tilde y} \right) = \left( {x - {\varphi _x}\left( {t,0} \right),y - {\varphi _y}\left( {t,0} \right)} \right)}}
\\ \nonumber
&& - \Delta x{\left. {\frac{\partial }{{\partial \tilde x}}{n_{t - i}^{ref}}\left( {\tilde x,\tilde y} \right)} \right|_{\left( {\tilde x,\tilde y} \right) = \left( {x - {\varphi _x}\left( {t,t - i} \right),y - {\varphi _y}\left( {t,t - i} \right)} \right)}}
\\ \nonumber
&& - \Delta y{\left. {\frac{\partial }{{\partial \tilde y}}{n_{t - i}^{ref}}\left( {\tilde x,\tilde y} \right)} \right|_{\left( {\tilde x,\tilde y} \right) = \left( {x - {\varphi _x}\left( {t,t - i} \right),y - {\varphi _y}\left( {t,t - i} \right)} \right)}}
\end{IEEEeqnarray}

The MC-prediction error of $f_t$ using $f_{t-i}$ as a reference frame was formulated in (\ref{eq:MC-prediction error}) as
\begin{equation}
\label{eq:MC-prediction error - appendix}
{e_{t|t - i}}\left( {x,y} \right) = {f_t}\left( {x,y} \right) - {{\hat f}_t}\left( {x,y\left| {{f_{t - i}^{ref}},\hat \varphi \left( {t,t - i\left| {{f_t},{f_{t - i}}} \right.} \right)} \right.} \right)
\end{equation}

Setting (\ref{eq:frame decomposition with translational MV}) and (\ref{eq:MC-prediction - approximated motion estimation model}) into (\ref{eq:MC-prediction error - appendix}) yields
\begin{IEEEeqnarray}{rCl}
\label{eq:MC-prediction - available frame error expression}
{e_{t|t - i}}\left( {x,y} \right) & = & \Delta x{\left. {\frac{\partial }{{\partial \tilde x}}v\left( {\tilde x,\tilde y} \right)} \right|_{\left( {\tilde x,\tilde y} \right) = \left( {x - {\varphi _x}\left( {t,0} \right),y - {\varphi _y}\left( {t,0} \right)} \right)}}
\\ \nonumber
&& + \Delta y{\left. {\frac{\partial }{{\partial \tilde y}}v\left( {\tilde x,\tilde y} \right)} \right|_{\left( {\tilde x,\tilde y} \right) = \left( {x - {\varphi _x}\left( {t,0} \right),y - {\varphi _y}\left( {t,0} \right)} \right)}}
\\ \nonumber
&& + \Delta x{\left. {\frac{\partial }{{\partial \tilde x}}{n_{t - i}^{ref}}\left( {\tilde x,\tilde y} \right)} \right|_{\left( {\tilde x,\tilde y} \right) = \left( {x - {\varphi _x}\left( {t,t - i} \right),y - {\varphi _y}\left( {t,t - i} \right)} \right)}}
\\ \nonumber
&& + \Delta y{\left. {\frac{\partial }{{\partial \tilde y}}{n_{t - i}^{ref}}\left( {\tilde x,\tilde y} \right)} \right|_{\left( {\tilde x,\tilde y} \right) = \left( {x - {\varphi _x}\left( {t,t - i} \right),y - {\varphi _y}\left( {t,t - i} \right)} \right)}}
\\ \nonumber
&& + {n_t}\left( {x,y} \right)
\\ \nonumber
&& - {n_{t - i}^{ref}}\left( {x - {\varphi _x}\left( {t,t - i} \right),y - {\varphi _y}\left( {t,t - i} \right)} \right)
\end{IEEEeqnarray}
We simplify the last expression by defining the motion-compensated noise difference, denoted as $\Delta {n_{t_2,t_1}}$ for $t_1 < t_2$:
\begin{IEEEeqnarray}{rCl}
\label{eq:MC-prediction - motion-compensated noise difference - definition - appendix}
&& \Delta {n_{t_2,t_1}}\left( {x,y} \right) \equiv {n_{t_2}}\left( {x,y} \right) - {n_{t_1}^{ref}}\left( {x - {\varphi _x}\left( {t_2,t_1} \right),y - {\varphi _y}\left( {t_2,t_1} \right)} \right)
\end{IEEEeqnarray}
Let us calculate $\Delta {n_{t_2,t_1}}$ for ${t_2} - {t_1} \le {L}$. Using $n_t$'s definition in (\ref{eq:accumulated noise definition}), and the corresponding $n_t^{ref}$ definition, we get 
\begin{IEEEeqnarray}{rCl}
\label{eq:MC-prediction - motion-compensated noise difference - calculation}
\Delta {n_{t_2,t_1}}\left( {x,y} \right) & = & w_{t_2} \left( {x,y} \right) + \sum\limits_{j = {t_2} - {L} + 1}^{{t_2}} {{q_j}\left( {x - {\varphi _x}\left( {{t_2},j} \right),y - {\varphi _y}\left( {{t_2},j} \right)} \right)}
\\ \nonumber
&& - w_{t_1}^{ref}\left( {x - {\varphi _x}\left( {t_2,t_1} \right),y - {\varphi _y}\left( {t_2,t_1} \right)} \right) 
\\ \nonumber
&& - \sum\limits_{h = {t_1} - {L} + 1}^{{t_1}} {q_h}\left( x - {\varphi _x}\left( {{t_2},{t_1}} \right) - {\varphi _x}\left( {{t_1},h} \right), y - {\varphi _y}\left( {{t_2},{t_1}} \right) - {\varphi _y}\left( {{t_1},h} \right) \right)
\\ \nonumber
& = & w_{t_2} \left( {x,y} \right) - w_{t_1}^{ref}\left( {x - {\varphi _x}\left( {t_2,t_1} \right),y - {\varphi _y}\left( {t_2,t_1} \right)} \right)
\\ \nonumber
&& + \sum\limits_{j = {t_1} + 1}^{{t_2}} {{q_j}\left( {x - {\varphi _x}\left( {{t_2},j} \right),y - {\varphi _y}\left( {{t_2},j} \right)} \right)}
\\ \nonumber
&& - \sum\limits_{h = {t_1} - {L} + 1}^{{t_2} - {L}} {{q_h}\left( {x - {\varphi _x}\left( {{t_2},h} \right),y - {\varphi _y}\left( {{t_2},h} \right)} \right)}
\end{IEEEeqnarray}
From $q_t$ and $w_t$'s independence property we get
\begin{IEEEeqnarray}{rCl}
\label{eq:MC-prediction - motion-compensated noise difference - autocorrelation}
{R_{\Delta {n_{{t_2},{t_1}}}}}\left( {k,l} \right) & = & \left[ {2\sigma _q^2\cdot\left( {{t_2} - {t_1}} \right) + \sigma _{w _{t_1}}^2 + \sigma _{w _{t_2}}^2 } \right] \cdot \delta \left( {k,l} \right)
\\ \nonumber
& = & \left[ {2\sigma _q^2\cdot\left( {{t_2} - {t_1}} \right) + 2\sigma _{w,basic}^2 + \sigma _{w,compression}^2} \right] \cdot \delta \left( {k,l} \right)
\end{IEEEeqnarray}

Let us calculate the autocorrelation of the error
\begin{IEEEeqnarray}{rCl}
\label{eq:MC-prediction - available frame error autocorrelation - definition}
{R_{{e_i}}}\left( {k,l} \right) = E\left\{ {{e_{t|t - i}}\left( {x,y} \right) \cdot {e_{t|t - i}}\left( {x + k,y + l} \right)} \right\}
\end{IEEEeqnarray}
In order to calculate $R_{{e_i}}$, we first analyze relations among the elements in $e_{t|t - i}$'s expression (\ref{eq:MC-prediction - available frame error expression}).
Recall that $\Delta x$ and $\Delta y$ are zero-mean independent variables; hence, $E\left\{ {\Delta x\Delta y} \right\} = 0$. Moreover, $\Delta x$ and $\Delta y$ are independent with $v$ and $n_j$ for any $j$; therefore, they are also independent with derivatives of $v$ and $n_j$, and the following relations hold
\begin{IEEEeqnarray}{rCl}
\label{eq:MC-prediction - available frame error autocorrelation - independent variables calculations}
\label{eq:MC-prediction - available frame error autocorrelation - independent variables calculations}
&& E\left\{ {{{\left( {\Delta x\frac{\partial }{{\partial \tilde x}}v\left( {\tilde x,\tilde y} \right)} \right)}^2}} \right\} = E\left\{ {\Delta {x^2}} \right\} \cdot E\left\{ {{{\left( {\frac{\partial }{{\partial \tilde x}}v\left( {\tilde x,\tilde y} \right)} \right)}^2}} \right\}
\\ \nonumber
&& E\left\{ {{{\left( {\Delta x\frac{\partial }{{\partial \tilde x}}{n_j}\left( {\tilde x,\tilde y} \right)} \right)}^2}} \right\} = E\left\{ {\Delta {x^2}} \right\} \cdot E\left\{ {{{\left( {\frac{\partial }{{\partial \tilde x}}{n_j}\left( {\tilde x,\tilde y} \right)} \right)}^2}} \right\}
\\ \nonumber
&& E\left\{ {\Delta x \cdot \frac{\partial }{{\partial \tilde x}}v\left( {\tilde x,\tilde y} \right) \cdot \Delta {n_{t,t - i}}\left( {x,y} \right)} \right\} = E\left\{ {\Delta x} \right\} \cdot E\left\{ {\frac{\partial }{{\partial \tilde x}}v\left( {\tilde x,\tilde y} \right)} \right\} \cdot E\left\{ {\Delta {n_{t,t - i}}\left( {x,y} \right)} \right\} = 0
\\ \nonumber
&& E\left\{ {\Delta x \cdot \frac{\partial }{{\partial \tilde x}}{n_{t - i}}\left( {\tilde x,\tilde y} \right) \cdot \Delta {n_{t,t - i}}\left( {x,y} \right)} \right\} = E\left\{ {\Delta x} \right\} \cdot E\left\{ {\frac{\partial }{{\partial \tilde x}}{n_{t - i}}\left( {\tilde x,\tilde y} \right) \cdot \Delta {n_{t,t - i}}\left( {x,y} \right)} \right\} = 0
\end{IEEEeqnarray}
${n_j}\left( {x,y} \right)$ is zero-mean and independent with $v\left( {x,y} \right)$, we utilize this in the following calculation.
\begin{IEEEeqnarray}{rCl}
\label{eq:MC-prediction - available frame error autocorrelation - independence of derivatives}
&& E\left\{ {\frac{\partial }{{\partial \tilde x}}v\left( {\tilde x,\tilde y} \right) \cdot \frac{\partial }{{\partial \tilde x}}{n_j}\left( {\tilde x,\tilde y} \right)} \right\} 
\\ \nonumber
&& \qquad {} \approx E\left\{ {\left( {\frac{{v\left( {\tilde x + {\varepsilon _v},\tilde y} \right) - v\left( {\tilde x,\tilde y} \right)}}{{{\varepsilon _v}}}} \right) \cdot \left( {\frac{{{n_j}\left( {\tilde x + {\varepsilon _n},\tilde y} \right) - {n_j}\left( {\tilde x,\tilde y} \right)}}{{{\varepsilon _n}}}} \right)} \right\}
\\ \nonumber
&& \qquad {} = E\left\{ {\frac{{v\left( {\tilde x + {\varepsilon _v},\tilde y} \right) - v\left( {\tilde x,\tilde y} \right)}}{{{\varepsilon _v}}}} \right\} \cdot E\left\{ {\frac{{{n_j}\left( {\tilde x + {\varepsilon _n},\tilde y} \right) - {n_j}\left( {\tilde x,\tilde y} \right)}}{{{\varepsilon _n}}}} \right\}
\\ \nonumber
&& \qquad {} = E\left\{ {\frac{{v\left( {\tilde x + {\varepsilon _v},\tilde y} \right) - v\left( {\tilde x,\tilde y} \right)}}{{{\varepsilon _v}}}} \right\} \cdot 0 = 0
\end{IEEEeqnarray}
Where $\varepsilon _v$ and $\varepsilon _n$ are very small. (\ref{eq:MC-prediction - available frame error autocorrelation - independent variables calculations})-(\ref{eq:MC-prediction - available frame error autocorrelation - independence of derivatives}) hold for $y$ by replacing $x$ with $y$.

Let us return to $R_{e_i}$'s calculation started in (\ref{eq:MC-prediction - available frame error autocorrelation - definition}). We define $\sigma _{\Delta x}^2 \equiv E\left\{ {\Delta {x^2}} \right\}$ and $\sigma _{\Delta y}^2 \equiv E\left\{ {\Delta {y^2}} \right\}$. Additionally, we define the following notation for the autocorrelation of the derivative of a function $f$:
\begin{IEEEeqnarray}{rCl}
\label{eq:notation for the autocorrelation of the derivative of a function}
&& autocorr\left\{ {{{\left. {\frac{\partial }{{\partial \tilde x}}f\left( {\tilde x,\tilde y} \right)} \right|}_{\left( {\tilde x,\tilde y} \right) = \left( {x,y} \right)}}} \right\}\left( {k,l} \right) \equiv 
\\ \nonumber
&& \qquad {} E\left\{ {{{\left. {\frac{\partial }{{\partial \tilde x}}f\left( {\tilde x,\tilde y} \right)} \right|}_{\left( {\tilde x,\tilde y} \right) = \left( {x,y} \right)}} \cdot {{\left. {\frac{\partial }{{\partial \tilde x}}f\left( {\tilde x,\tilde y} \right)} \right|}_{\left( {\tilde x,\tilde y} \right) = \left( {x + k,y + l} \right)}}} \right\}
\end{IEEEeqnarray}
We use (\ref{eq:MC-prediction - available frame error autocorrelation - independent variables calculations})-(\ref{eq:notation for the autocorrelation of the derivative of a function}) to get
\begin{IEEEeqnarray}{rCl}
\label{eq:MC-prediction - available frame error autocorrelation}
{R_{{e_i}}}\left( {k,l} \right) & = & \sigma _{\Delta x}^2\cdot autocorr\left\{ {{{\left. {\frac{\partial }{{\partial \tilde x}}v\left( {\tilde x,\tilde y} \right)} \right|}_{\left( {\tilde x,\tilde y} \right) = \left( {\tilde x _0,\tilde y _0} \right)}}} \right\}\left( {k,l} \right)
\\ \nonumber
&& + \sigma _{\Delta y}^2\cdot autocorr\left\{ {{{\left. {\frac{\partial }{{\partial \tilde y}}v\left( {\tilde x,\tilde y} \right)} \right|}_{\left( {\tilde x,\tilde y} \right) = \left( {\tilde x _0,\tilde y _0} \right)}}} \right\}\left( {k,l} \right)
\\ \nonumber
&& + \sigma _{\Delta x}^2\cdot autocorr\left\{ {{{\left. {\frac{\partial }{{\partial \tilde x}}{n_{t - i}^{ref}}\left( {\tilde x,\tilde y} \right)} \right|}_{\left( {\tilde x,\tilde y} \right) = \left( {\tilde x _i,\tilde y _i} \right)}}} \right\}\left( {k,l} \right)
\\ \nonumber
&& + \sigma _{\Delta y}^2\cdot autocorr\left\{ {{{\left. {\frac{\partial }{{\partial \tilde y}}{n_{t - i}^{ref}}\left( {\tilde x,\tilde y} \right)} \right|}_{\left( {\tilde x,\tilde y} \right) = \left( {\tilde x _i,\tilde y _i} \right)}}} \right\}\left( {k,l} \right)
\\ \nonumber
&& + E\left\{ {\Delta {n_{t,t - i}}\left( {x,y} \right) \cdot \Delta {n_{t,t - i}}\left( {x + k,y + l} \right)} \right\}
\end{IEEEeqnarray}
where $\left( {\tilde x _0,\tilde y _0} \right) = \left( {x - {\varphi _x}\left( {t,0} \right),y - {\varphi _y}\left( {t,0} \right)} \right)$ and $\left( {\tilde x _i,\tilde y _i} \right) = \left( {x - {\varphi _x}\left( {t,t - i} \right),y - {\varphi _y}\left( {t,t - i} \right)} \right)$.

The discrete derivative approximation is
\begin{IEEEeqnarray}{rCl}
\label{eq:Discrete derivation approximation}
{\left. {\frac{\partial }{{\partial \tilde x}}f\left( {\tilde x,\tilde y} \right)} \right|_{\left( {\tilde x,\tilde y} \right) = \left( {x,y} \right)}} \approx f\left( {x + 1,y} \right) - f\left( {x,y} \right).
\end{IEEEeqnarray}
By using this approximation, we represent the derivative autocorrelation as function of $f$'s autocorrelation, $R_f$:
\begin{IEEEeqnarray}{rCl}
\label{eq:Derivative autocorrelation using the discrete derivation approximation}
&& autocorr\left\{ {{{\left. {\frac{\partial }{{\partial \tilde x}}f\left( {\tilde x,\tilde y} \right)} \right|}_{\left( {\tilde x,\tilde y} \right) = \left( {x,y} \right)}}} \right\}\left( {k,l} \right)
\\ \nonumber
&& \qquad {} = E\left\{ \left[ {f\left( {x + 1,y} \right) - f\left( {x,y} \right)} \right] \cdot \left[ {f\left( {x + 1 + k,y + l} \right) - f\left( {x + k,y + l} \right)} \right] \right\}
\\ \nonumber 
&& \qquad {} = 2{R_f}\left( {k,l} \right) - {R_f}\left( {k - 1,l} \right) - {R_f}\left( {k + 1,l} \right)
\end{IEEEeqnarray}
We use (\ref{eq:Derivative autocorrelation using the discrete derivation approximation}) to eliminate the derivative operators in (\ref{eq:MC-prediction - available frame error autocorrelation}):
\begin{IEEEeqnarray}{rCl}
\label{eq:MC-prediction - available frame error autocorrelation - as function of other aotucorrelation functions}
{R_{{e_i}}}\left( {k,l} \right) & = & 2\left[ {\sigma _{\Delta x}^2 + \sigma _{\Delta y}^2} \right] \cdot \left[ {{R_v}\left( {k,l} \right) + {R_{{n_{t - i}^{ref}}}}\left( {k,l} \right)} \right]
\\ \nonumber
&& - \sigma _{\Delta x}^2\cdot\left[ {{R_v}\left( {k - 1,l} \right) + {R_v}\left( {k + 1,l} \right)} \right]
\\ \nonumber
&& - \sigma _{\Delta x}^2\cdot\left[ {{R_{{n_{t - i}^{ref}}}}\left( {k - 1,l} \right) + {R_{{n_{t - i}^{ref}}}}\left( {k + 1,l} \right)} \right]
\\ \nonumber
&& - \sigma _{\Delta y}^2\cdot\left[ {{R_v}\left( {k,l - 1} \right) + {R_v}\left( {k,l + 1} \right)} \right]
\\ \nonumber
&& - \sigma _{\Delta y}^2\cdot\left[ {{R_{{n_{t - i}^{ref}}}}\left( {k,l - 1} \right) + {R_{{n_{t - i}^{ref}}}}\left( {k,l + 1} \right)} \right]
\\ \nonumber
&& + {R_{\Delta {n_{t,t - i}}}}\left( {k,l} \right)
\end{IEEEeqnarray}
Substituting autocorrelation expressions from (\ref{eq:pure image autocorrelation}), (\ref{eq:accumulated noise autocorrelation}) and (\ref{eq:MC-prediction - motion-compensated noise difference - autocorrelation}) into (\ref{eq:MC-prediction - available frame error autocorrelation - as function of other aotucorrelation functions}) yields
\begin{IEEEeqnarray}{rCl}
\label{eq:Available frame error autocorrelation - explicit}
{R_{{e_i}}}\left( {k,l} \right) & = & 2\left[ {\sigma _{\Delta x}^2 + \sigma _{\Delta y}^2} \right]\cdot\left[ {\sigma _v^2\cdot\rho _v^{\left| k \right| + \left| l \right|} + \left( {L \sigma _q^2 + \sigma _{w,ref}^2} \right)\cdot\delta \left( {k,l} \right)} \right]
\\ \nonumber
&& - \sigma _{\Delta x}^2\sigma _v^2{\rho_v ^{\left| l \right|}} \cdot \left[ {{\rho_v ^{\left| {k - 1} \right|}} + {\rho_v^{\left| {k + 1} \right|}}} \right]
\\ \nonumber
&& - \sigma _{\Delta x}^2\left[ {L\sigma _q^2 + \sigma _{w,ref}^2} \right]\cdot\left[ {\delta \left( {k - 1,l} \right) + \delta \left( {k + 1,l} \right)} \right]
\\ \nonumber
&& - \sigma _{\Delta y}^2\sigma _v^2{\rho_v^{\left| k \right|}} \cdot \left[ {{\rho_v ^{\left| {l - 1} \right|}} + {\rho_v^{\left| {l + 1} \right|}}} \right]
\\ \nonumber
&& - \sigma _{\Delta y}^2\left[ {L\sigma _q^2 + \sigma _{w,ref}^2} \right]\cdot\left[ {\delta \left( {k,l - 1} \right) + \delta \left( {k,l + 1} \right)} \right]
\\ \nonumber
&& + \left[ {2i\sigma _q^2 + \sigma _{w,current}^2 + \sigma _{w,ref}^2} \right]\cdot\delta \left( {k,l} \right)
\end{IEEEeqnarray}

\newpage
\section{Autocorrelation Calculation for Prediction of an Absent Frame}
\label{appendix_sec:MC Prediction of an Absent Frame}
Here we calculate the autocorrelation of the MC-prediction residual for the case of an available frame that was presented in section \ref{subsec:MC-Prediction of an Absent Frame}.
Recall our definitions from section \ref{subsec:MC-Prediction of an Absent Frame} for the backward prediction:
\begin{IEEEeqnarray}{rCl}
\label{eq:MC-prediction - absent frame - f0 prediction}
\nonumber && {{\hat f}_j}\left( {x,y\left| {{f_0},} \right.\hat \varphi \left( {j,0\left| {{f_0},{f_D}} \right.} \right)} \right) = {f_0}\left( {x - {{\hat \varphi }_x}\left( {j,0\left| {{f_0},{f_D}} \right.} \right),y - {{\hat \varphi }_y}\left( {j,0\left| {{f_0},{f_D}} \right.} \right)} \right).
\\
\end{IEEEeqnarray}
and the forward prediction:
\begin{IEEEeqnarray}{rCl}
\label{eq:MC-prediction - absent frame - fD prediction}
\nonumber && {{\hat f}_j}\left( {x,y\left| {{f_D},} \right.\hat \varphi \left( {D,j\left| {{f_0},{f_D}} \right.} \right)} \right) = {f_D}\left( {x + {{\hat \varphi }_x}\left( {D,j\left| {{f_0},{f_D}} \right.} \right),y + {{\hat \varphi }_y}\left( {D,j\left| {{f_0},{f_D}} \right.} \right)} \right).
\\
\end{IEEEeqnarray}
Similar to (\ref{eq:MC-prediction - approximated motion estimation model}), we get
\begin{IEEEeqnarray}{rCl}
\label{eq:MC-prediction - absent frame - approximated motion estimation model - backward}
{{\hat f}_j}\left( {x,y\left| {{f_0},} \right.\hat \varphi \left( {j,0\left| {{f_0},{f_D}} \right.} \right)} \right) & = & v\left( {x - {\varphi _x}\left( {j,0} \right),y - {\varphi _y}\left( {j,0} \right)} \right)
\\ \nonumber
&& - \Delta x_0^{abs}{\left. {\frac{\partial }{{\partial \tilde x}}v\left( {\tilde x,\tilde y} \right)} \right|_{\left( {\tilde x,\tilde y} \right) = \left( {x - {\varphi _x}\left( {j,0} \right),y - {\varphi _y}\left( {j,0} \right)} \right)}}
\\ \nonumber
&& - \Delta y_0^{abs}{\left. {\frac{\partial }{{\partial \tilde y}}v\left( {\tilde x,\tilde y} \right)} \right|_{\left( {\tilde x,\tilde y} \right) = \left( {x - {\varphi _x}\left( {j,0} \right),y - {\varphi _y}\left( {j,0} \right)} \right)}}
\\ \nonumber
&& + {n_0}\left( {x - {\varphi _x}\left( {j,0} \right),y - {\varphi _y}\left( {j,0} \right)} \right)
\\ \nonumber
&& - \Delta x_0^{abs}{\left. {\frac{\partial }{{\partial \tilde x}}{n_0}\left( {\tilde x,\tilde y} \right)} \right|_{\left( {\tilde x,\tilde y} \right) = \left( {x - {\varphi _x}\left( {j,0} \right),y - {\varphi _y}\left( {j,0} \right)} \right)}}
\\ \nonumber
&& - \Delta y_0^{abs}{\left. {\frac{\partial }{{\partial \tilde y}}{n_0}\left( {\tilde x,\tilde y} \right)} \right|_{\left( {\tilde x,\tilde y} \right) = \left( {x - {\varphi _x}\left( {j,0} \right),y - {\varphi _y}\left( {j,0} \right)} \right)}}
\end{IEEEeqnarray}
and
\begin{IEEEeqnarray}{rCl}
\label{eq:MC-prediction - absent frame - approximated motion estimation model - forward}
{{\hat f}_j}\left( {x,y\left| {{f_D},} \right.\hat \varphi \left( {D,j\left| {{f_0},{f_D}} \right.} \right)} \right) & = & v\left( {x - {\varphi _x}\left( {j,0} \right),y - {\varphi _y}\left( {j,0} \right)} \right)
\\ \nonumber
&& + \Delta x_D^{abs}{\left. {\frac{\partial }{{\partial \tilde x}}v\left( {\tilde x,\tilde y} \right)} \right|_{\left( {\tilde x,\tilde y} \right) = \left( {x - {\varphi _x}\left( {j,0} \right),y - {\varphi _y}\left( {j,0} \right)} \right)}}
\\ \nonumber
&& + \Delta y_D^{abs}{\left. {\frac{\partial }{{\partial \tilde y}}v\left( {\tilde x,\tilde y} \right)} \right|_{\left( {\tilde x,\tilde y} \right) = \left( {x - {\varphi _x}\left( {j,0} \right),y - {\varphi _y}\left( {j,0} \right)} \right)}}
\\ \nonumber
&& + {n_D}\left( {x + {\varphi _x}\left( {D,j} \right),y + {\varphi _y}\left( {D,j} \right)} \right)
\\ \nonumber
&& + \Delta x_D^{abs}{\left. {\frac{\partial }{{\partial \tilde x}}{n_D}\left( {\tilde x,\tilde y} \right)} \right|_{\left( {\tilde x,\tilde y} \right) = \left( {x + {\varphi _x}\left( {D,j} \right),y + {\varphi _y}\left( {D,j} \right)} \right)}}
\\ \nonumber
&& + \Delta y_D^{abs}{\left. {\frac{\partial }{{\partial \tilde y}}{n_D}\left( {\tilde x,\tilde y} \right)} \right|_{\left( {\tilde x,\tilde y} \right) = \left( {x + {\varphi _x}\left( {D,j} \right),y + {\varphi _y}\left( {D,j} \right)} \right)}}
\end{IEEEeqnarray}

The final prediction was defined in (\ref{eq:MC-prediction - absent frame - final prediction}) as
\begin{IEEEeqnarray}{rCl}
\label{eq:MC-prediction - absent frame - final prediction - appendix}
\hat f_j^{final}\left( {x,y\left| {{f_0},} \right.{f_D}} \right) & = & \theta \cdot {{\hat f}_j}\left( {x,y\left| {{f_0},} \right.\hat \varphi \left( {j,0\left| {{f_0},{f_D}} \right.} \right)} \right) 
\\ \nonumber
&& + \left[ {1 - \theta } \right] \cdot {{\hat f}_j}\left( {x,y\left| {{f_D},} \right.\hat \varphi \left( {D,j\left| {{f_0},{f_D}} \right.} \right)} \right)
\end{IEEEeqnarray}
Setting (\ref{eq:MC-prediction - absent frame - approximated motion estimation model - backward}) and (\ref{eq:MC-prediction - absent frame - approximated motion estimation model - forward}) into (\ref{eq:MC-prediction - absent frame - final prediction - appendix}) yields
\begin{IEEEeqnarray}{rCl}
\label{eq:MC-prediction - absent frame - final prediction - explicit}
\hat f_j^{final}\left( {x,y\left| {{f_0},} \right.{f_D}} \right) & = & v\left( {x - {\varphi _x}\left( {j,0} \right),y - {\varphi _y}\left( {j,0} \right)} \right)
\\ \nonumber
&& - \left[ {\theta  \cdot \Delta x_0^{abs} - \left( {1 - \theta } \right) \cdot \Delta x_D^{abs}} \right] \cdot {\left. {\frac{\partial }{{\partial \tilde x}}v\left( {\tilde x,\tilde y} \right)} \right|_{\left( {\tilde x,\tilde y} \right) = \left( {x - {\varphi _x}\left( {j,0} \right),y - {\varphi _y}\left( {j,0} \right)} \right)}}
\\ \nonumber
&& - \left[ {\theta  \cdot \Delta y_0^{abs} - \left( {1 - \theta } \right) \cdot \Delta y_D^{abs}} \right] \cdot {\left. {\frac{\partial }{{\partial \tilde y}}v\left( {\tilde x,\tilde y} \right)} \right|_{\left( {\tilde x,\tilde y} \right) = \left( {x - {\varphi _x}\left( {j,0} \right),y - {\varphi _y}\left( {j,0} \right)} \right)}}
\\ \nonumber
&& + \theta \cdot\Bigg[ {n_0}\left( {x - {\varphi _x}\left( {j,0} \right),y - {\varphi _y}\left( {j,0} \right)} \right) 
\\ \nonumber
&& \qquad \left. {} - \Delta x_0^{abs}{{\left. {\frac{\partial }{{\partial \tilde x}}{n_0}\left( {\tilde x,\tilde y} \right)} \right|}_{\left( {\tilde x,\tilde y} \right) = \left( {x - {\varphi _x}\left( {j,0} \right),y - {\varphi _y}\left( {j,0} \right)} \right)}} \right.\\ \nonumber
&& \qquad \left. {} - \Delta y_0^{abs}{{\left. {\frac{\partial }{{\partial \tilde y}}{n_0}\left( {\tilde x,\tilde y} \right)} \right|}_{\left( {\tilde x,\tilde y} \right) = \left( {x - {\varphi _x}\left( {j,0} \right),y - {\varphi _y}\left( {j,0} \right)} \right)}} \right]
\\ \nonumber
&& + \left( {1 - \theta } \right)\cdot\Bigg[ {n_D}\left( {x + {\varphi _x}\left( {D,j} \right),y + {\varphi _y}\left( {D,j} \right)} \right) 
\\ \nonumber
&& \qquad\qquad \left. {} + \Delta x_D^{abs}{{\left. {\frac{\partial }{{\partial \tilde x}}{n_D}\left( {\tilde x,\tilde y} \right)} \right|}_{\left( {\tilde x,\tilde y} \right) = \left( {x + {\varphi _x}\left( {D,j} \right),y + {\varphi _y}\left( {D,j} \right)} \right)}} 
\right.\\ \nonumber
&& \qquad\qquad \left. {} + \Delta y_D^{abs}{{\left. {\frac{\partial }{{\partial \tilde y}}{n_D}\left( {\tilde x,\tilde y} \right)} \right|}_{\left( {\tilde x,\tilde y} \right) = \left( {x + {\varphi _x}\left( {D,j} \right),y + {\varphi _y}\left( {D,j} \right)} \right)}} \right]
\end{IEEEeqnarray}

The prediction error was defined in (\ref{eq:MC-prediction - absent frame error expression}) as
\begin{IEEEeqnarray}{rCl}
\label{eq:MC-prediction - absent frame error expression - appendix}
e_{j|0,D}^{absent}\left( {x,y} \right) & = & {f_j}\left( {x,y} \right) - \hat f_j^{final}\left( {x,y\left| {{f_0},} \right.{f_D}} \right)
\end{IEEEeqnarray}
We develop the last error expression as follows:
\begin{IEEEeqnarray}{rCl}
\label{eq:MC-prediction - absent frame error expression - detailed}
e_{j|0,D}^{absent}\left( {x,y} \right) & = & {f_j}\left( {x,y} \right) - \hat f_j^{final}\left( {x,y\left| {{f_0},} \right.{f_D}} \right)
\\ \nonumber
& = & {n_j}\left( {x,y} \right) + \left[ {\theta \Delta x_0^{abs} - \left( {1 - \theta } \right) \Delta x_D^{abs}} \right] \cdot {\left. {\frac{\partial }{{\partial \tilde x}}v\left( {\tilde x,\tilde y} \right)} \right|_{\left( {\tilde x,\tilde y} \right) = \left( {x - {\varphi _x}\left( {j,0} \right),y - {\varphi _y}\left( {j,0} \right)} \right)}}
\\ \nonumber
&& + \left[ {\theta \Delta y_0^{abs} - \left( {1 - \theta } \right) \Delta y_D^{abs}} \right] \cdot {\left. {\frac{\partial }{{\partial \tilde y}}v\left( {\tilde x,\tilde y} \right)} \right|_{\left( {\tilde x,\tilde y} \right) = \left( {x - {\varphi _x}\left( {j,0} \right),y - {\varphi _y}\left( {j,0} \right)} \right)}}
\\ \nonumber
&& - \theta \Bigg[ {n_0}\left( {x - {\varphi _x}\left( {j,0} \right),y - {\varphi _y}\left( {j,0} \right)} \right) 
\\ \nonumber
&& \qquad \left. {} - \Delta x_0^{abs}{{\left. {\frac{\partial }{{\partial \tilde x}}{n_0}\left( {\tilde x,\tilde y} \right)} \right|}_{\left( {\tilde x,\tilde y} \right) = \left( {x - {\varphi _x}\left( {j,0} \right),y - {\varphi _y}\left( {j,0} \right)} \right)}} 
\right.\\ \nonumber
&& \qquad \left. {} - \Delta y_0^{abs}{{\left. {\frac{\partial }{{\partial \tilde y}}{n_0}\left( {\tilde x,\tilde y} \right)} \right|}_{\left( {\tilde x,\tilde y} \right) = \left( {x - {\varphi _x}\left( {j,0} \right),y - {\varphi _y}\left( {j,0} \right)} \right)}} \right]
\\ \nonumber
&& - \left( {1 - \theta } \right)\Bigg[ {n_D}\left( {x + {\varphi _x}\left( {D,j} \right),y + {\varphi _y}\left( {D,j} \right)} \right) 
\\ \nonumber
&& \qquad\qquad \left. {} + \Delta x_D^{abs}{{\left. {\frac{\partial }{{\partial \tilde x}}{n_D}\left( {\tilde x,\tilde y} \right)} \right|}_{\left( {\tilde x,\tilde y} \right) = \left( {x + {\varphi _x}\left( {D,j} \right),y + {\varphi _y}\left( {D,j} \right)} \right)}} 
\right.\\ \nonumber
&& \qquad\qquad \left. {} + \Delta y_D^{abs}{{\left. {\frac{\partial }{{\partial \tilde y}}{n_D}\left( {\tilde x,\tilde y} \right)} \right|}_{\left( {\tilde x,\tilde y} \right) = \left( {x + {\varphi _x}\left( {D,j} \right),y + {\varphi _y}\left( {D,j} \right)} \right)}} \right]
\\ \nonumber
& = & \theta  \left[ {{n_j}\left( {x,y} \right) - {n_0}\left( {x - {\varphi _x}\left( {j,0} \right),y - {\varphi _y}\left( {j,0} \right)} \right)} \right]
\\ \nonumber
&& - \left( {1 - \theta } \right) \left[ {{n_D}\left( {x + {\varphi _x}\left( {D,j} \right),y + {\varphi _y}\left( {D,j} \right)} \right) - {n_j}\left( {x,y} \right)} \right]
\\ \nonumber
&& + \left[ {\theta \Delta x_0^{abs} - \left( {1 - \theta } \right) \Delta x_D^{abs}} \right] \cdot {\left. {\frac{\partial }{{\partial \tilde x}}v\left( {\tilde x,\tilde y} \right)} \right|_{\left( {\tilde x,\tilde y} \right) = \left( {x - {\varphi _x}\left( {j,0} \right),y - {\varphi _y}\left( {j,0} \right)} \right)}}
\\ \nonumber
&& + \left[ {\theta \Delta y_0^{abs} - \left( {1 - \theta } \right) \Delta y_D^{abs}} \right] \cdot {\left. {\frac{\partial }{{\partial \tilde y}}v\left( {\tilde x,\tilde y} \right)} \right|_{\left( {\tilde x,\tilde y} \right) = \left( {x - {\varphi _x}\left( {j,0} \right),y - {\varphi _y}\left( {j,0} \right)} \right)}}
\\ \nonumber
&& + \theta  \left[ \Delta x_0^{abs}{\left. {\frac{\partial }{{\partial \tilde x}}{n_0}\left( {\tilde x,\tilde y} \right)} \right|_{\left( {\tilde x,\tilde y} \right) = \left( {x - {\varphi _x}\left( {j,0} \right),y - {\varphi _y}\left( {j,0} \right)} \right)}} \right.
\\ \nonumber
&& \qquad {} \left.+ \Delta y_0^{abs}{\left. {\frac{\partial }{{\partial \tilde y}}{n_0}\left( {\tilde x,\tilde y} \right)} \right|_{\left( {\tilde x,\tilde y} \right) = \left( {x - {\varphi _x}\left( {j,0} \right),y - {\varphi _y}\left( {j,0} \right)} \right)}} \right]
\\ \nonumber
&& - \left( {1 - \theta } \right) \left[ 
\Delta x_D^{abs}{\left. {\frac{\partial }{{\partial \tilde x}}{n_D}\left( {\tilde x,\tilde y} \right)} \right|_{\left( {\tilde x,\tilde y} \right) = \left( {x + {\varphi _x}\left( {D,j} \right),y + {\varphi _y}\left( {D,j} \right)} \right)}} \right.
\\ \nonumber
&& \qquad\qquad {} \left. + \Delta y_D^{abs}{\left. {\frac{\partial }{{\partial \tilde y}}{n_D}\left( {\tilde x,\tilde y} \right)} \right|_{\left( {\tilde x,\tilde y} \right) = \left( {x + {\varphi _x}\left( {D,j} \right),y + {\varphi _y}\left( {D,j} \right)} \right)}} \right]
\end{IEEEeqnarray}
From $\Delta n$'s definition in (\ref{eq:MC-prediction - motion-compensated noise difference - definition}) we get
\begin{IEEEeqnarray}{rCl}
\label{eq:MC-prediction - absent frame - delta n for substitution}
&& \Delta {n_{j,0}}\left( {x,y} \right) = {n_j}\left( {x,y} \right) - {n_0}\left( {x - {\varphi _x}\left( {j,0} \right),y - {\varphi _y}\left( {j,0} \right)} \right)
\\ \nonumber
\\ \nonumber
&& \Delta {n_{D,j}}\left( {x + {\varphi _x}\left( {D,j} \right),y + {\varphi _y}\left( {D,j} \right)} \right) = {n_D}\left( {x + {\varphi _x}\left( {D,j} \right),y + {\varphi _y}\left( {D,j} \right)} \right) - {n_j}\left( {x,y} \right)
\end{IEEEeqnarray}
Inserting (\ref{eq:MC-prediction - absent frame - delta n for substitution}) into (\ref{eq:MC-prediction - absent frame error expression - detailed}) yields
\begin{IEEEeqnarray}{rCl}
\label{eq:MC-prediction - absent frame error expression}
e_{j|0,D}^{absent}\left( {x,y} \right) & = & \theta  \cdot \Delta {n_{j,0}}\left( {x,y} \right) 
\\ \nonumber
&& - \left( {1 - \theta } \right) \cdot \Delta {n_{D,j}}\left( {x + {\varphi _x}\left( {D,j} \right),y + {\varphi _y}\left( {D,j} \right)} \right)
\\ \nonumber
&& + \left[ {\theta  \cdot \Delta x_0^{abs} - \left( {1 - \theta } \right) \cdot \Delta x_D^{abs}} \right] \cdot {\left. {\frac{\partial }{{\partial \tilde x}}v\left( {\tilde x,\tilde y} \right)} \right|_{\left( {\tilde x,\tilde y} \right) = \left( {x - {\varphi _x}\left( {j,0} \right),y - {\varphi _y}\left( {j,0} \right)} \right)}}
\\ \nonumber
&& + \left[ {\theta  \cdot \Delta y_0^{abs} - \left( {1 - \theta } \right) \cdot \Delta y_D^{abs}} \right] \cdot {\left. {\frac{\partial }{{\partial \tilde y}}v\left( {\tilde x,\tilde y} \right)} \right|_{\left( {\tilde x,\tilde y} \right) = \left( {x - {\varphi _x}\left( {j,0} \right),y - {\varphi _y}\left( {j,0} \right)} \right)}}
\\ \nonumber
&& + \theta \cdot\left[ \Delta x_0^{abs}{{\left. {\frac{\partial }{{\partial \tilde x}}{n_0}\left( {\tilde x,\tilde y} \right)} \right|}_{\left( {\tilde x,\tilde y} \right) = \left( {x - {\varphi _x}\left( {j,0} \right),y - {\varphi _y}\left( {j,0} \right)} \right)}} \right.\\ \nonumber
&& \qquad \left. {} + \Delta y_0^{abs}{{\left. {\frac{\partial }{{\partial \tilde y}}{n_0}\left( {\tilde x,\tilde y} \right)} \right|}_{\left( {\tilde x,\tilde y} \right) = \left( {x - {\varphi _x}\left( {j,0} \right),y - {\varphi _y}\left( {j,0} \right)} \right)}} \right]
\\ \nonumber
&& - \left( {1 - \theta } \right)\cdot\left[ \Delta x_D^{abs}{{\left. {\frac{\partial }{{\partial \tilde x}}{n_D}\left( {\tilde x,\tilde y} \right)} \right|}_{\left( {\tilde x,\tilde y} \right) = \left( {x + {\varphi _x}\left( {D,j} \right),y + {\varphi _y}\left( {D,j} \right)} \right)}} \right.\\ \nonumber
&& \qquad\qquad \left. {} + \Delta y_D^{abs}{{\left. {\frac{\partial }{{\partial \tilde y}}{n_D}\left( {\tilde x,\tilde y} \right)} \right|}_{\left( {\tilde x,\tilde y} \right) = \left( {x + {\varphi _x}\left( {D,j} \right),y + {\varphi _y}\left( {D,j} \right)} \right)}} \right]
\end{IEEEeqnarray}

Let us derive an expression for the error autocorrelation.
\begin{IEEEeqnarray}{rCl}
\label{eq:MC-prediction - absent frame - error autocorrelation}
&& {R_{e_{j|0,D}^{absent}}}\left( {k,l} \right) = E\left\{ {e_{j|0,D}^{absent}\left( {x,y} \right) \cdot e_{j|0,D}^{absent}\left( {x + k,y + l} \right)} \right\}
\\ \nonumber
&& \qquad {} = {\theta ^2} \cdot {R_{\Delta {n_{j,0}}}}\left( {k,l} \right) + {\left( {1 - \theta } \right)^2} \cdot {R_{\Delta {n_{D,j}}}}\left( {k,l} \right) + 
\\ \nonumber
&& \qquad {} + \sigma _{\Delta {x^{abs}}}^2 \left[ {{\theta ^2} + {{\left( {1 - \theta } \right)}^2}} \right] \cdot autocorr\left\{ {{{\left. {\frac{\partial }{{\partial \tilde x}}v\left( {\tilde x,\tilde y} \right)} \right|}_{\left( {\tilde x,\tilde y} \right) = \left( {x - {\varphi _x}\left( {j,0} \right),y - {\varphi _y}\left( {j,0} \right)} \right)}}} \right\}
\\ \nonumber
&& \qquad {} + \sigma _{\Delta {y^{abs}}}^2 \left[ {{\theta ^2} + {{\left( {1 - \theta } \right)}^2}} \right] \cdot autocorr\left\{ {{{\left. {\frac{\partial }{{\partial \tilde y}}v\left( {\tilde x,\tilde y} \right)} \right|}_{\left( {\tilde x,\tilde y} \right) = \left( {x - {\varphi _x}\left( {j,0} \right),y - {\varphi _y}\left( {j,0} \right)} \right)}}} \right\}
\\ \nonumber
&& \qquad {} + {\theta ^2}\sigma _{\Delta {x^{abs}}}^2 \cdot autocorr\left\{ {{{\left. {\frac{\partial }{{\partial \tilde x}}{n_0}\left( {\tilde x,\tilde y} \right)} \right|}_{\left( {\tilde x,\tilde y} \right) = \left( {x - {\varphi _x}\left( {j,0} \right),y - {\varphi _y}\left( {j,0} \right)} \right)}}} \right\}
\\ \nonumber
&& \qquad {} + {\theta ^2}\sigma _{\Delta {y^{abs}}}^2 \cdot autocorr\left\{ {{{\left. {\frac{\partial }{{\partial \tilde y}}{n_0}\left( {\tilde x,\tilde y} \right)} \right|}_{\left( {\tilde x,\tilde y} \right) = \left( {x - {\varphi _x}\left( {j,0} \right),y - {\varphi _y}\left( {j,0} \right)} \right)}}} \right\}
\\ \nonumber
&& \qquad {} + {\left( {1 - \theta } \right)^2}\sigma _{\Delta {x^{abs}}}^2 \cdot autocorr\left\{ {{{\left. {\frac{\partial }{{\partial \tilde x}}{n_D}\left( {\tilde x,\tilde y} \right)} \right|}_{\left( {\tilde x,\tilde y} \right) = \left( {x - {\varphi _x}\left( {D,j} \right),y - {\varphi _y}\left( {D,j} \right)} \right)}}} \right\}
\\ \nonumber
&& \qquad {} + {\left( {1 - \theta } \right)^2}\sigma _{\Delta {y^{abs}}}^2 \cdot autocorr\left\{ {{{\left. {\frac{\partial }{{\partial \tilde y}}{n_D}\left( {\tilde x,\tilde y} \right)} \right|}_{\left( {\tilde x,\tilde y} \right) = \left( {x - {\varphi _x}\left( {D,j} \right),y - {\varphi _y}\left( {D,j} \right)} \right)}}} \right\}
\\ \nonumber
&& \qquad {} + \theta \left( {1 - \theta } \right) E\left\{ \Delta {n_{j,0}}\left( {x,y} \right) \cdot \Delta {n_{D,j}}\left( {x + {\varphi _x}\left( {D,j} \right) + k,y + {\varphi _y}\left( {D,j} \right) + l} \right) \right\}
\\ \nonumber
&& \qquad {} + \theta \left( {1 - \theta } \right) E\left\{ \Delta {n_{j,0}}\left( {x + k,y + l} \right) \cdot \Delta {n_{D,j}}\left( {x + {\varphi _x}\left( {D,j} \right),y + {\varphi _y}\left( {D,j} \right)} \right) \right\}
\end{IEEEeqnarray}
The cross-correlation between ${\Delta {n_{j,0}}}$ and ${\Delta {n_{D,j}}}$ is
\begin{IEEEeqnarray}{rCl}
\label{eq:MC-prediction - absent frame - delta noise cross correlation}
&& E\left\{ \Delta {n_{j,0}}\left( {x,y} \right) \cdot \Delta {n_{D,j}}\left( {x + {\varphi _x}\left( {D,j} \right) + k,y + {\varphi _y}\left( {D,j} \right) + l} \right) \right\} = 0
\end{IEEEeqnarray}
and
\begin{IEEEeqnarray}{rCl}
\label{eq:MC-prediction - absent frame - delta noise cross correlation 2}
&& E\left\{ \Delta {n_{j,0}}\left( {x + k,y + l} \right) \cdot \Delta {n_{D,j}}\left( {x + {\varphi _x}\left( {D,j} \right),y + {\varphi _y}\left( {D,j} \right)} \right) \right\} = 0
\end{IEEEeqnarray}
Setting (\ref{eq:Derivative autocorrelation using the discrete derivation approximation}), (\ref{eq:MC-prediction - absent frame - delta noise cross correlation}) and (\ref{eq:MC-prediction - absent frame - delta noise cross correlation 2}) into (\ref{eq:MC-prediction - absent frame - error autocorrelation}) results in
\begin{IEEEeqnarray}{rCl}
\label{eq:MC-prediction - absent frame - error autocorrelation as function of other autocorrelation functions}
{R_{e_{j|0,D}^{absent}}}\left( {k,l} \right) & = & {\theta ^2} \cdot {R_{\Delta {n_{j,0}}}}\left( {k,l} \right) + {\left( {1 - \theta } \right)^2} \cdot {R_{\Delta {n_{D,j}}}}\left( {k,l} \right) + 
\\ \nonumber
&& + \sigma _{\Delta {x^{abs}}}^2 \cdot \left[ {{\theta ^2} + {{\left( {1 - \theta } \right)}^2}} \right] \cdot \left[ {2{R_v}\left( {k,l} \right) - {R_v}\left( {k - 1,l} \right) - {R_v}\left( {k + 1,l} \right)} \right]
\\ \nonumber
&& + \sigma _{\Delta {y^{abs}}}^2 \cdot \left[ {{\theta ^2} + {{\left( {1 - \theta } \right)}^2}} \right] \cdot \left[ {2{R_v}\left( {k,l} \right) - {R_v}\left( {k,l - 1} \right) - {R_v}\left( {k,l + 1} \right)} \right]
\\ \nonumber
&& + {\theta ^2}\sigma _{\Delta {x^{abs}}}^2 \cdot \left[ {2{R_{{n_0}}}\left( {k,l} \right) - {R_{{n_0}}}\left( {k - 1,l} \right) - {R_{{n_0}}}\left( {k + 1,l} \right)} \right.
\\ \nonumber
&& \qquad\qquad\qquad {} + \left. {2{R_{{n_D}}}\left( {k,l} \right) - {R_{{n_D}}}\left( {k - 1,l} \right) - {R_{{n_D}}}\left( {k + 1,l} \right)} \right]
\\ \nonumber
&& + {\theta ^2}\sigma _{\Delta {y^{abs}}}^2 \cdot \left[ {2{R_{{n_0}}}\left( {k,l} \right) - {R_{{n_0}}}\left( {k,l - 1} \right) - {R_{{n_0}}}\left( {k,l + 1} \right)} \right.
\\ \nonumber
&& \qquad\qquad\qquad {} + \left. {2{R_{{n_D}}}\left( {k,l} \right) - {R_{{n_D}}}\left( {k,l - 1} \right) - {R_{{n_D}}}\left( {k,l + 1} \right)} \right]
\end{IEEEeqnarray}
According to (\ref{eq:accumulated noise autocorrelation}), $R_{{n_t}}$ is time-invariant; i.e., it does not depend on $t$. Hence, $R_{{n_0}} \equiv R_{{n_D}}$ and (\ref{eq:MC-prediction - absent frame - error autocorrelation as function of other autocorrelation functions}) is simplified to
\begin{IEEEeqnarray}{rCl}
\label{eq:MC-prediction - absent frame - error autocorrelation as function of other autocorrelation functions - simplified form}
{R_{e_{j|0,D}^{absent}}}\left( {k,l} \right) & = & {\theta ^2} \cdot {R_{\Delta {n_{j,0}}}}\left( {k,l} \right) + {\left( {1 - \theta } \right)^2} \cdot {R_{\Delta {n_{D,j}}}}\left( {k,l} \right) + 
\\ \nonumber
&& + \sigma _{\Delta {x^{abs}}}^2 \cdot \left[ {{\theta ^2} + {{\left( {1 - \theta } \right)}^2}} \right]  
\\ \nonumber
&& \qquad\qquad {} \times \left[ {2{R_v}\left( {k,l} \right) - {R_v}\left( {k - 1,l} \right) - {R_v}\left( {k + 1,l} \right)} \right. 
\\ \nonumber
&& \qquad\qquad\qquad {} \left. + {2{R_{{n_0}}}\left( {k,l} \right) - {R_{{n_0}}}\left( {k - 1,l} \right) - {R_{{n_0}}}\left( {k + 1,l} \right)} 
 \right]
\\ \nonumber
&& + \sigma _{\Delta {y^{abs}}}^2 \cdot \left[ {{\theta ^2} + {{\left( {1 - \theta } \right)}^2}} \right] 
\\ \nonumber
&& \qquad\qquad {} \times \left[ {2{R_v}\left( {k,l} \right) - {R_v}\left( {k,l - 1} \right) - {R_v}\left( {k,l + 1} \right)} \right. 
\\ \nonumber
&& \qquad\qquad\qquad {} \left. + {2{R_{{n_0}}}\left( {k,l} \right) - {R_{{n_0}}}\left( {k,l - 1} \right) - {R_{{n_0}}}\left( {k,l + 1} \right)} \right]
\end{IEEEeqnarray}

%
%

\newpage



\bibliographystyle{IEEEtran}
\bibliography{IEEEabrv,mc_paper_refs}
%

%
%




\end{document}